\documentclass{article}
\usepackage[utf8]{inputenc}
\usepackage{amsmath}
\usepackage{amssymb}
\usepackage{amsfonts}
\usepackage{esvect}
\usepackage{cite}
\usepackage{physics}
\usepackage{mathrsfs}
\usepackage{authblk}
\usepackage{geometry}
\usepackage{abstract}
\usepackage{xcolor}
\newtheorem{proposition}{Proposition}

\title{\bf Decomposition of enveloping algebras of simple Lie algebras and their related polynomial algebras   }
\author{\large Rutwig Campoamor-Stursberg$^{1}$\footnote{rutwig@ucm.es}, 
Ian Marquette$^{2}$ \footnote{i.marquette@uq.edu.au} }
\affil{$^{1}$ Instituto de Matem\'atica Interdisciplinar and Dpto. Geometr\'{\i}a y
Topolog\'{\i}a,
UCM,\\E-28040 Madrid, Spain}
\affil{$^{2}$ School of Mathematics and Physics, The University of Queensland \\ Brisbane, QLD 4072, Australia}

\usepackage[colorlinks=true,urlcolor=black]{hyperref}

\begin{document}

\maketitle
\begin{abstract}
The decomposition problem of the enveloping algebra of a simple Lie algebra is reconsidered combining both the analytical and the algebraic approach, showing its relation with the internal labelling problem with respect to a nilpotent subalgebra. A lower bound for the number of generators of the commutant as well as the maximal Abelian subalgebra are obtained. The decomposition problem for the exceptional Lie algebra $G_2$ is completely solved.

\end{abstract}

\section{Introduction}

Decomposition theorems for enveloping algebras of (simple) Lie algebras constitute a classical result within the structure theory, and have found extensive application in representation theory \cite{Kos}. For physical problems, these techniques have passed unnoticed for a long time, until the systematic study of superintegrable and quasi-solvable systems has shown how the detailed analysis of enveloping algebras, as well as its analytical counterpart in terms of differential operators can be of much use in solving these problems \cite{Ovs,Tur,Post,Lat}. This has motivated various approaches to polynomial algebras, both from the perspective of rings of functions as well as from the purely algebraic formalism (see \cite{Frei,Leto,Poli,Ian,cam21} and references therein). Polynomial algebras have attracted attention on their own right, in particular, the quadratic case \cite{Poli}, showing to what extent different properties of Lie algebras, such as the Jacobi identity or the Poincar\' e--Birkhoff--Witt bases are preserved, providing a new scope for building other types of algebraic structures. Polynomial algebras of this type have also found applications within the context of Gr\"{o}bner bases and adjacent areas \cite{Li}, the generalization of space-time conformal algebras \cite{Jar}, as well as the description of symmetries and orthogonal polynomials related to quantum models \cite{Post,Lat}. This shows the diversity of problems in which polynomial algebras emerge naturally. Here, we intend to point out the connection between polynomial algebras and some classical problems of simple Lie algebras and their enveloping algebra. 

\medskip
In this work we reconsider the decomposition problem of enveloping algebras of simple Lie algebras combining the purely algebraic approach with an analytical interpretation. More precisely, we show that finding the commutant $C_{\mathcal{U}(\mathfrak{s})}(P)$ of the highest weight vector for the adjoint representation is computationally equivalent to determine the subgroup scalars for a distinguished nilpotent Lie algebra, namely the nilradical $\mathfrak{n}$ of the Borel subalgebra of $\mathfrak{s}$. This allows us to estimate a lower bound for the number of polynomials in the enveloping algebra required to span the commutant, as well as to determine the dimension of its maximal Abelian subalgebra. The case of the three classical rank-two simple Lie algebras, known from the literature (see \cite{Fla,Bur1} and references therein) are revisited from this perspective. Further, we solve the decomposition problem for the exceptional Lie algebra $G_2$, which is considerably involved from the computational point of view. It is shown that the polynomials giving rise to the decomposition of the enveloping algebra determine a $17$-dimensional polynomial algebra admitting several non-Abelian subalgebras. In the general context, it is shown that the commutant $C_{\mathcal{U}(\mathfrak{s})}(P)$ always admits a non-Abelian polynomial subalgebra $\mathcal{A}$ spanned by polynomials forming an integrity basis for the system of partial differential equations (PDEs) associated to the reduction chain $\mathfrak{n}\subset \mathfrak{s}$. We finish with some comments on how the procedure can be extended to reduction chains of semisimple Lie algebras not related to the decomposition of enveloping algebras.

\section{Presentation of semisimple Lie algebras}

One of the most convenient descriptions of complex semisimple Lie algebras is given in terms of the so-called Serre relations, that allow to reconstruct the Lie algebra from its root system and the Dynkin diagram \cite{serr}.

\medskip
Let $\mathfrak{s}$ be a complex semisimple Lie algebra of rank $\ell$, $\left\{H_1,\dots ,H_{\ell}\right\}$ a basis of a Cartan subalgebra $\mathfrak{h}$, $\mathcal{R}$ the corresponding root system and $\Delta=\left\{\alpha_1,\dots ,\alpha_{\ell}\right\}$ a basis of simple roots. Recall that the  Cartan integers are thus given by 
\begin{equation*}
n(i,j)=\langle \alpha_j,\alpha_i\rangle = \frac{2 \left(\alpha_j,\alpha_i\right)}{\left(\alpha_i,\alpha_i\right)},
\end{equation*}
where $\left(\alpha_i,\alpha_j\right)=\kappa\left(H_i,H_j\right)$ is defined in terms of the nondegenerate Killing form $\kappa$. Then we can always find $3\ell$  generators $H_i,X_i,Y_i$ satisfying the relations 
\begin{equation}\label{serrel}
\begin{split}
\left[H_i,H_j\right]& =0,\quad  \left[H_i,X_j\right]=n(i,j) X_j,\quad \left[H_i,Y_j\right]=-n(i,j)Y_j,\\
\left[X_i,Y_j\right]& =\delta_i^j H_i,\quad  {\rm ad}(X_i)^{1-n(i,j)}X_j=0, \quad  {\rm ad}(Y_i)^{1-n(i,j)}Y_j=0\; (i\neq j).
\end{split}
\end{equation}
Clearly, the generators $X_i$ are associated to the simple roots $\alpha_i$, while the $Y_i$ correspond to the opposite roots (see e.g. \cite{serr}). 

\medskip
To any linear form $\alpha\in \mathfrak{h}^{\ast}$ we associate the weight space $\mathfrak{s}_{\alpha}$ defined by 
\begin{equation*}
\mathfrak{s}_{\alpha}=\left\{ X\in\mathfrak{s}\;|\; [H,X]=\alpha(H) X,\quad H\in\mathfrak{h}\right\},
\end{equation*}
Clearly $\mathfrak{h}=\mathfrak{s}^{0}$ and the $\alpha$ such that $\mathfrak{s}^{\alpha}\neq 0$ correspond to the roots in $\mathcal{R}$. In terms of the weight spaces, the Lie algebra $\mathfrak{s}$ can be decomposed as 
\begin{equation}\label{CW}
\mathfrak{s}=\mathfrak{h}\oplus \sum_{\alpha\in \mathcal{R}} \mathfrak{s}_{\alpha}. 
\end{equation}
This actually induces a grading in $\mathfrak{s}$, as for any $\alpha_i,\alpha_j\in R$ the relations (\ref{serrel}) imply that $\left[\mathfrak{s}^{\alpha_i},\mathfrak{s}^{\alpha_j}\right]\subset \mathfrak{s}^{\alpha_i+\alpha_j}$.  
\medskip
\noindent Let thus $ \mathcal{ U}(\mathfrak{s})$ be the universal enveloping algebra of $\mathfrak{s}$ . For any fixed positive integer $p$, we denote by $\mathcal{U}_{(p)}(\mathfrak{g})$ the subspace generated by the monomials $X_1^{a_1}\dots X_n^{a_n}$ satisfying the constraint $a_1+a_2+\dots +a_n\leq p$, where $n=\dim\mathfrak{s}$. Using this relation, we say that an element $P\in \mathcal{U}(\mathfrak{s})$  is of degree $d$ if  $d={\rm inf}\left\{k\;|\; P\in \mathcal{U}_{(k)}(\mathfrak{s})\right\}$. As the enveloping algebra is naturally filtered, we have the relation   
\begin{equation}
\mathcal{U}_{(0)}(\mathfrak{s})=\mathbb{C},\; \mathcal{U}_{(p)}(\mathfrak{s})\mathcal{U}_{(q)}(\mathfrak{s})\subset \mathcal{U}_{(p
+q)}(\mathfrak{s}),\quad p,q\geq 0.\label{fil1}
\end{equation}
A direct consequence of the filtration is that each $\mathcal{U}_{(p)}(\mathfrak{s})$ is a finite-dimensional representation of $\mathfrak{s}$, hence the enveloping algebra $\mathcal{U}(\mathfrak{s})$ is a sum of finite-dimensional representations of the simple Lie algebra $\mathfrak{s}$. 

\medskip
\noindent From the general structure theory (see e.g. \cite{Dix,AA}) it is easily seen that the adjoint action of $\mathfrak{s}$ on $\mathcal{U}(\mathfrak{s})$ (respectively, the symmetric algebra $S(\mathfrak{s})$) is given by 
\begin{equation}\label{adja}
\begin{array}[c]{rl}
P\in \mathcal{U}(\mathfrak{s}) \mapsto & P.X_i:= \left[X_i,P\right]= X_i P-P X_i\in\mathcal{U}(\mathfrak{s}),\\
P\left(x_1,\dots ,x_n\right)\in S(\mathfrak{g})\mapsto &\displaystyle  \widehat{X}_i(P)=C_{ij}^{k} x_k \frac{\partial}{\partial x_j}\in S(\mathfrak{g}),\\
\end{array},
\end{equation}
where $\left\{X_1,\dots ,X_n\right\}$ is a basis of $\mathfrak{s}$. A linear isomorphism $\Lambda:S(\mathfrak{s})\rightarrow \mathcal{U}(\mathfrak{s})$ that commutes with the adjoint action is easily obtained through the symmetrization map
\begin{equation}\label{syma}
\Lambda\left(x_{j_1}\dots x_{j_p}\right)=\frac{1}{p!} \sum_{\sigma\in{\bf S}_{p}} X_{j_{\sigma(1)}}\dots X_{j_{\sigma(p)}},
\end{equation}
where ${\bf S}(p)$ denotes the symmetric group of order $p!$. For the space $S^{(p)}(\mathfrak{s})$ of homogeneous polynomials having degree $p$, the relation $\mathcal{U}^{(p)}(\mathfrak{s})=\Lambda\left(S^{(p)}(\mathfrak{s})\right)$ further allows us to decompose the subspace $\mathcal{U}_{(p)}(\mathfrak{g})$ as a direct sum of homogeneous elements: $\mathcal{U}_{(p)}(\mathfrak{s})=\sum_{k=0}^{p} \mathcal{U}^{(k)}(\mathfrak{s})$. Thus for any $P\in \mathcal{U}_{(p)}(\mathfrak{s}), Q\in\mathcal{U}_{(q)}(\mathfrak{s})$ the commutator satisfies 
\begin{equation*}
\left[ P,Q\right]\in \mathcal{U}_{(p+q-1)}(\mathfrak{s}).
\end{equation*}
Further, as a direct consequence of the Poincar\'e--Birkhoff--Witt theorem, we have the dimension formula 
\begin{equation}\label{difo}
\dim \mathcal{U}^{(p)}(\mathfrak{s})=\dim \frac{\mathcal{U}_{(p)}(\mathfrak{s})}{\mathcal{U}_{(p-1)}(\mathfrak{s})}=\dim S^{(p)}(\mathfrak{s})=\left(\begin{array}[c]{c}
\dim\mathfrak{s}+p-1\\
p\end{array}\right).
\end{equation}
Invariant polynomials of $\mathfrak{s}$ are defined as the centre of $\mathcal{U}(\mathfrak{s})$:
\begin{equation}\label{INVS1}
Z\left(\mathcal{U}(\mathfrak{s})\right) = \left\{ P\in\mathcal{U}(\mathfrak{s})\; |\;\left[\mathfrak{s},P\right]=0\right\}.
\end{equation}
These elements can be identified with the (polynomial) solutions of the differential operators in(\ref{adja}) (see \cite{Dix,AA} for details). 
 
\medskip
\noindent  The commutant $C_{\mathcal{U}(\mathfrak{s})}(P)$ of an element $P\in\mathcal{U}(\mathfrak{s})$ is defined as the centralizer of $P$ in the enveloping algebra, that is, the set of elements in $\mathcal{U}(\mathfrak{g})$ that commute with $P$ 
\begin{equation}
C_{\mathcal{U}(\mathfrak{s})}(P)=\left\{ Q\in\mathcal{U}(\mathfrak{s})\; |\; [P,Q]=0\right\}.\label{comm}
\end{equation}
 As $\mathfrak{s}$ is semisimple, the commutant is finitely generated, a property that allows us to find an integrity basis for it.

\subsection{Decomposition of enveloping algebras and commutants}

\noindent As mentioned before, the adjoint action (\ref{adja}) implies that the universal enveloping algebra $\mathcal{U}(\mathfrak{s})$ can be seen as a representation of $\mathfrak{s}$, which is moreover completely reducible whenever the Lie algebra is semisimple \cite{Dix}. We therefore have a decomposition
\begin{equation}\label{udek}
\mathcal{U}(\mathfrak{s})=\bigoplus_{k=0}^{\infty}V_{k},
\end{equation}
 where $V_k=\mathcal{U}(\mathfrak{s}).v_{\lambda_k}$ is an irreducible component spanned by a highest vector $v_{\lambda_k}$ with weight $\lambda_k\in\mathfrak{h}^{\ast}$ for the extended adjoint action, i.e.,  
\begin{equation*}
\left[H_i,v_{\lambda_k}\right]=\lambda_k(H_i)v_{\lambda_k},\; H_i\in\mathfrak{h},\quad 
\left[X_{i},v_{\lambda_k}\right]=0,\; X_i\in\mathfrak{s}_{\alpha_i},\; \alpha_i\in \Delta. 
\end{equation*}
Therefore, finding the components in the decomposition (\ref{udek}) amounts to compute the elements in the enveloping algebra $\mathcal{U}(\mathfrak{s})$ that commute with the generators associated to the positive roots of $\mathfrak{s}$. Due to the Serre relations (\ref{serrel}), it suffices to consider the elements in $\mathcal{U}(\mathfrak{s})$ that commute with the generators $X_i$ associated to simple roots $\alpha_i\in \Delta$, as any other elements of $\mathfrak{n}$ are obtained by commutators. In this context, we observe that the generators $X_i$ associated to simple roots generate a nilpotent Lie algebra $\mathfrak{n}$ (see equation (\ref{serrel})).\footnote{This algebra  actually corresponds to the nilradical $\mathfrak{n}$ of the Borel subalgebra $\mathfrak{b}(\Delta)$ of $\mathfrak{s}$ (the maximal solvable subalgebra of $\mathfrak{s}$)}, so that the decomposition problem  corresponds to find the centralizer of $\mathfrak{n}$ in $\mathcal{U}(\mathfrak{s})$:
\begin{equation}
\mathcal{C}_{\mathcal{U}(\mathfrak{s})}(\mathfrak{n})=\left\{ P\in \mathcal{U}(\mathfrak{s})\; | \; [X,P]=0,\quad X\in \mathfrak{n}\right\}
\end{equation}

\medskip

By the algebraic properties of enveloping algebras, we can always find a finite number of polynomials $\left\{P_1,\dots ,P_s\right\}$ such that, as vector space, $\mathcal{C}_{\mathcal{U}(\mathfrak{s})}(\mathfrak{n})$ is spanned by 
\begin{equation}\label{bas1}
P_1^{a_1}P_2^{a_2}\dots P_s^{a_s},\quad a_i\in\mathbb{N}\cup {0},
\end{equation}
and where the coefficients $a_i$ are possibly constrained by some relation. It should be observed that these polynomials are not necessarily algebraically independent, but merely linearly independent, and that they generate a non-Abelian polynomial algebra. This implies in practice that suitable dependent polynomials must be added to a functionally independent set of polynomials in order to obtain a basis of the type (\ref{bas1}). With respect to a given Cartan subalgebra $\mathfrak{h}$, if the weight of each $P_i$ is given by $\left(\lambda_1^i,\dots ,\lambda_\ell^i\right)$, the weight of an element (\ref{bas1}) is given by 
\begin{equation}
\omega_i= \sum_{i=1}^{s} a_i \left(\lambda_1^i,\dots ,\lambda_\ell^i\right).
\end{equation}
Any element $P\in \mathcal{C}_{\mathcal{U}(\mathfrak{s})}(\mathfrak{n})$ can thus be labeled using its degree $d$ as polynomial and its weight $\mu=\left(\lambda_1,\dots ,\lambda_\ell\right)$ with respect to the (fixed) Cartan subalgebra.
For given values of $a_i$, the element in (\ref{bas1}) generates an irreducible representation of $\mathfrak{s}$ of dimension $d_{a_1,\dots a_s}$. In order to fulfill the decomposition (\ref{udek}), the identity 
\begin{equation}\label{dek1}
\dim\mathcal{U}^{(p)}=\sum_{(a_1,\dots ,a_s)\in \Phi_p}d_{a_1,\dots a_s},\quad 
\Phi_p=\left\{(a_1,\dots ,a_s)\;|\; \sum_{k=1}^{s}a_k{\rm deg} P_k=p\right\}\quad {\rm mod}\; \mathcal{S},\quad p\geq 1
\end{equation}
must be satisfied, where $\mathcal{S}$ denotes the set of relations that constrain the coefficients $a_i$. 

\medskip
As the commutant $\mathcal{C}_{\mathcal{U}(\mathfrak{s})}(\mathfrak{n})$ involves generators that do not belong to the subalgebra $\mathfrak{n}$, in addition to the invariants of $\mathfrak{n}$ there will be additional polynomials depending on the variables of $\mathfrak{s}$, in particular the Casimir operators of the simple Lie algebra $\mathfrak{s}$. We can suppose without loss of generalization that for the generic element (\ref{bas1}), the (primitive) Casimir operators $I_1,\dots I_\ell$ of $\mathfrak{s}$ satisfy  
\begin{equation}
P_{s+1-\ell}=I_1,\dots ,P_s=I_\ell.
\end{equation}
Defining the set  
\begin{equation}
\mathcal{B}=\left\{P_1^{a_1}P_2^{a_2}\dots P_{s-\ell}^{a_{s-\ell}},\quad a_i\in\mathbb{N}\cup {0}\right\} ,
\end{equation}
it follows easily that $\mathcal{B}$ forms a basis of the commutant $\mathcal{C}_{\mathcal{U}(\mathfrak{s})}(\mathfrak{n})$ seen as a free module over $\mathbb{C}\left[I_1,\dots ,I_\ell\right]$. This is an immediate consequence of the Schur lemma, as the Casimir operators act as scalar matrices on each irreducible $\mathfrak{s}$-representation. As the invariants of $\mathfrak{s}$ have zero highest weight (with respect to a Cartan subalgebra), the highest weight of the vectors in (\ref{bas1}) are determined by the polynomials which have a nonvanishing commutator with at least one generator of $\mathfrak{s}$ not belonging to the nilpotent subalgebra $\mathfrak{n}$. In particular, the generator $X_{\beta}$ corresponding to the highest root $\beta\in\mathcal{R}^+$ always has a nonzero weight, as it lies in the closure of the Weyl chamber. The Casimir operators of $\mathfrak{n}$, as they do not commute with all generators of $\mathfrak{s}$, also have a nonvanishing weight. 
In terms of representation theory, this implies that for any finite-dimensional irreducible complex representation $\varphi:\mathfrak{s}\rightarrow {\rm End}(V)$, the elements $\varphi(P)$ with $P\in\mathcal{B}$ span the space of highest weight vectors in the representation \cite{Fla}.

\section{Analytic properties of the commutant}
 
\medskip
\noindent Alternatively to the analysis of the universal enveloping algebra $\mathcal{U}(\mathfrak{s})$, we can use the differential operators in (\ref{adja}) to compute the elements in the commutant analytically. As $\mathfrak{n}$ is a subalgebra, this correspond to the so-called labelling problem with respect to the chain $\mathfrak{n}\subset\mathfrak{s}$ (see \cite{Ra,Pe,C97}
). 

\medskip
\noindent Starting from the embedding $\mathfrak{n}\subset\mathfrak{s}$, we extend an arbitrary  basis $\left\{X_{1},\ldots,X_{m}\right\}$ of the subalgebra $\mathfrak{n}$ to a basis $\left\{X_{1},\dots,X_{m},Y_1,\dots Y_{n-m}\right\}$ of $\mathfrak{s}$, such that the commutators of the subalgebra generators are given by  
\begin{equation}\label{suba1}
\left[ X_i,X_j\right]= C_{ij}^k X_k,\quad \left[ X_i,Y_p\right]= D_{ip}^k X_k +E_{ip}^q Y_q
\end{equation}
where $i,j,k\in\left\{1,\dots ,m\right\}$ and $p,q\in\left\{1,\dots ,n-m\right\}$. The operators that commute with the elements in $\mathfrak{n}$ hence correspond to the solutions of the system of PDEs 
\begin{equation}
\widehat{X}_{i}=-C_{ij}^{k}x_{k}\frac{\partial }{\partial x_{j}}-(D_{ip}^k x_k +E_{ip}^q y_q )\frac{\partial }{\partial y_{p}},\quad 1\leq i\leq m.
\label{Rep2}
\end{equation}
where $\left\{ x_{1},\ldots,x_{n},y_{1},\ldots,y_{n-m}\right\}$ are the coordinates in a dual basis of $\mathfrak{B}$. We observe that solutions $F$ to the system (\ref{Rep2}) such that $\displaystyle \frac{\partial F}{\partial y_{p}}=0$ for all $p\in\left\{1,\dots ,n-m\right\}$ correspond to the Casimir invariants of the subalgebra, while a genuine subgroup scalar must explicitly depend on the variables $\left\{  y_{1},\ldots,y_{n-m}\right\}$ \cite{C97}. Now the system (\ref{Rep2}) has exactly $n-r^{\prime}$ independent solutions, where $r^{\prime}$ denotes the rank of the $m\times n$ coefficient matrix of $\mathfrak{n}$ as subalgebra of $\mathfrak{s}$. Although an arbitrary integrity basis for the system (\ref{Rep2}) does not necessarily generate a polynomial algebra, so that the analytical approach provides at best a lower bound $d_0$ on the number of required polynomials to obtain the decomposition (\ref{udek}), it may be asked whether for any simple Lie algebra there exists at least an integrity basis for the system (\ref{Rep2}) such that forms a polynomial algebra. We observe that starting from such a subalgebra, the commutant $\mathcal{C}_{\mathcal{U}(\mathfrak{s})}(\mathfrak{n})$ can formally be constructed, adding those elements that are algebraically dependent (but linearly independent) required to satisfy the dimension condition (\ref{dek1}). 

\medskip
The following table gives the lower bound $d_0$ obtained analytically for the complex simple Lie algebras: 
\begin{equation}
\begin{array}{c|ccccccccc}
{\rm Type} & A_\ell & B_\ell,C_\ell & D_\ell & G_2 & F_4 & E_6 & E_7 & E_8   \\
d_0 & \displaystyle \frac{\ell(\ell+3)}{2} & \ell^2+\ell & \ell^2 & 8 & 28 & 42 & 70 & 128   
\end{array}
\end{equation}
Among these polynomials, the invariants of $\mathfrak{n}$ are included. As the subalgebra is nilpotent, we can always find a maximal set of functionally independent solutions formed by polynomials, i.e., an integrity basis \cite{Dix65}. In particular, its centre is generated by the generator associated to the highest root in $\mathcal{R}$, and the commutant contains exactly one linear polynomial. From these solutions, $\ell+\mathcal{N}(\mathfrak{n})$ correspond to the Casimir invariants of either $\mathfrak{s}$ or $\mathfrak{n}$,\footnote{Here we use the fact that the nilradical of the Borel subalgebra and $\mathfrak{s}$ can never have common Casimir invariants.} so that the number of available operators is given by $\chi=n-r^{\prime}-\ell-\mathcal{N}(\mathfrak{n})$. It can be easily shown (see e.g. \cite{Pe}) that $m=r^{\prime}$, which implies that $\chi= 2n_0$. It should however be noted that among these $2n_0$ solutions, at most $n_0$ correspond to operators that commute with each other \cite{Pe}. We conclude from this that the maximal number of operators that commute with the subalgebra $\mathfrak{n}$ and with each other is given by 
\begin{equation}\label{abel}
\xi=n_0+\ell+\mathcal{N}(\mathfrak{n})=\frac{\dim\mathfrak{s}-\dim\mathfrak{n}+\ell+\mathcal{N}(\mathfrak{n})}{2}.
\end{equation}
The remaining independent solutions of system (\ref{Rep2}) necessarily have some nontrivial commutator. Expressed in other words, we conclude that the commutant $\mathcal{C}_{\mathcal{U}(\mathfrak{s})}(\mathfrak{n})$ always contains a maximal Abelian subalgebra of dimension $\xi$.

\medskip As the generators of $\mathfrak{n}$ correspond to the positive roots of $\mathfrak{s}$, we can easily deduce the number of Casimir invariants of $\mathfrak{n}$ making use of the so-called Maurer--Cartan equations of the subalgebra \cite{C43}. From (\ref{CW}) we know that 
$\mathfrak{n}=\sum_{\alpha\in\mathcal{R}^+}s_\alpha$. Let $\omega_{\alpha}$ denote the invariant 1-form corresponding to the generator of the weight space $s_\alpha$. For any positive root, the Maurer--Cartan equations can be written formally as
\begin{equation}
{\rm d}\omega_{\alpha}= \sum_{\beta,\gamma\in\mathcal{R}^+} \lambda^{\beta,\gamma}\omega_{\beta}\wedge\omega_{\gamma},
\end{equation}
where $ \lambda^{\beta,\gamma}=0$ whenever $\beta+\gamma\neq \alpha$. In particular, if $\alpha_i\in\Delta$ is a simple root, it is obvious that the form is closed, i.e, ${\rm d}\omega_{\alpha_i}=0$, while for non-simple roots the relations ${\rm d}\omega_{\alpha}\neq 0$ holds. We now define a generic 2-form $\theta=\sum_{\alpha\in\mathcal{R}^+}{\rm d}\omega_\alpha$ and the index $j_0(\mathfrak{n})$ as the lowest natural number such that
\begin{equation}
\bigwedge^{j_0(\mathfrak{n})}\theta\neq 0,\quad \bigwedge^{j_0(\mathfrak{n})+1}\theta\equiv 0.
\end{equation}
Then the number $\mathcal{N}(\mathfrak{n})$ of invariants is given by $\mathcal{N}(\mathfrak{n})=\dim\mathfrak{n}-2j_0(\mathfrak{n})$ \cite{C43}.

\begin{proposition}
Let $\mathfrak{n}$ be the nilradical of the Borel subalgebra of a complex simple  Lie algebra $\mathfrak{s}$. 
\begin{enumerate}
\item If $\mathfrak{s}\simeq A_\ell$, $\dim\mathfrak{n}=\displaystyle\frac{\ell(\ell+1)}{2}$ and  $\mathcal{N}(\mathfrak{n})=\displaystyle\left[\frac{\ell+1}{2}\right]$ for $\ell\geq 1$. 

\item If $\mathfrak{s}\simeq B_\ell$, $\dim\mathfrak{n}=\displaystyle\ell^2$ and  $\mathcal{N}(\mathfrak{n})=\displaystyle \ell $ for $\ell\geq 2$. 

\item If $\mathfrak{s}\simeq C_\ell$, $\dim\mathfrak{n}=\displaystyle\ell^2$ and  $\mathcal{N}(\mathfrak{n})=\displaystyle \ell $ for $\ell\geq 3$. 

\item If $\mathfrak{s}\simeq D_\ell$, $\dim\mathfrak{n}=\displaystyle\ell^2-\ell$ and  
\begin{equation*}
\mathcal{N}(\mathfrak{n})=\left\{\begin{array}[c]{cl}
\ell & \ell=2p,\; p\geq 2\\
\ell-1 & \ell=2p+1,\; p\geq 2
\end{array}\right.
\end{equation*}

\item If $\mathfrak{s}$ is an exceptional Lie algebra of rank $\ell$, then $\mathcal{N}(\mathfrak{n})=\ell$.
\end{enumerate}
\end{proposition}

We prove the assertion for the type $A_\ell$, the argument being the same for the remaining types of simple classical algebras. The positive roots $\mathcal{R}^+$ of $A_\ell$ are given by 
\begin{equation}
\beta_{j,k}=\sum_{p=j}^{k} \alpha_p,\quad j,k=1,\dots \ell,\; j\leq k
\end{equation}
Let $\omega_\alpha$ denote the 1-forms. It is clear from the structure of the root system of $A_\ell$ that the Maurer--Cartan forms have the generic form 
\begin{equation}
{\rm d}\omega_{\beta_{j,k}}= \sum_{s=j}^{k} \lambda_{j,k}^{s}\omega_{\beta_{j,s}}\wedge\omega_{\beta_{s+1,k}},
\end{equation}
with coefficients $\lambda_{j,k}^{s}$ such that ${\rm d}\left({\rm d}\omega_{\beta_{j,k}}\right)=0$ holds. We separate the analysis according to the parity of $\ell$. Let $\ell=2p$ and consider the 2-form
\begin{equation}
\theta={\rm d}\omega_{\beta_{1,2p}}+\sum_{q=1}^{p-1}\;\sum_{s=0}^{p-q-\frac{1}{2}} {\rm d}\omega_{\beta_{2q,2q+2s+1}}+\sum_{q=1}^{p-1}\;\sum_{s=0}^{p-q-2} {\rm d}\omega_{\beta_{2q+1,2q+2s+2}}
\end{equation}
A routine computation shows that the form $\theta$ satisfies \begin{equation}
\bigwedge^{p^2}\theta\neq 0,\quad \bigwedge^{p^2+1}\theta\equiv 0,
\end{equation}
and that adding additional 2-forms ${\rm d}\beta_{j,k}$ does not increase the degree of $\theta$,
thus $j_0(\mathfrak{n})=p^2$. Therefore the number of Casimir invariants of $\mathcal{n}$ is given by 
\begin{equation}
\mathcal{N}(\mathfrak{n})=2p^2+p-2p^2=p =\displaystyle\left[\frac{2p+1}{2}\right]
\end{equation}
For odd $\ell=2p+1$, we consider the 2-form   
\begin{equation}
\theta={\rm d}\omega_{\beta_{1,2p}}+\sum_{q=1}^{p-1}\;\sum_{s=0}^{p-q-\frac{1}{2}} {\rm d}\omega_{\beta_{2q,2q+2s+1}}+\sum_{q=1}^{p-1}\;\sum_{s=0}^{p-q-1} {\rm d}\omega_{\beta_{2q+1,2q+2s+2}}
\end{equation}
satisfying $\displaystyle \bigwedge^{p^2+p}\theta\neq 0,\quad \bigwedge^{p^2+p+1}\theta\equiv 0$, and the addition of other 2-form does not increase the degree $j_0(\mathfrak{n})=p^2+p$, from which we deduce  that $\mathcal{N}(\mathfrak{n})=2p^2+3p+1-2p^2-2p=p+1 =\displaystyle\left[\frac{2p+2}{2}\right]$. 

\medskip

\section{Analysis of rank $2$ semisimple Lie algebras}  

\subsection{$D_2=so(4)$}

Let us first consider the semisimple but not simple Lie algebra $D_2=A_1^2$. In this case, the root system is reducible and indeed the disjoint union of the root systems corresponding to $A_1$. Choosing a basis adapted to the root system, the commutators are 
\begin{equation}
\begin{split}
[X_1,X_2]=2X_2 ,\quad [X_1,X_3]=-2X_3,\quad [X_2,X_3]=X_1,\\
[X_4,X_5]=2 X_5 ,\quad [X_4,X_6]=-2 X_6,\quad [X_5,X_6]=X_4.
\end{split}
\end{equation}
As expected from the fact that $\mathfrak{n}=\mathbb{R}\langle X_2,X_4\rangle$  is a decomposable algebra, the commutant is also decomposable. There are four algebraically independent polynomials in $\mathcal{C}_{\mathcal{U}(\mathfrak{s})}(\mathfrak{n})$, given by $X_1,X_4$ themselves, as well as the two Casimir operators $I_{21}=X_1^2+2(X_2X_3+X_3X_2)$, $I_{22}=X_4^2+2(X_5X_6+X_6X_5)$ of $D_2$. These elements form an integrity basis for the solutions of the system (\ref{Rep2}), and generate the (Abelian) commutant. Hence any element in $\mathcal{C}_{\mathcal{U}(\mathfrak{s})}(\mathfrak{n})$ can be written as 
\begin{equation}
P= X_1^{a_1}X_4^{a_2}I_{21}^{a_3}I_{22}^{a_4},\quad a_i\in\mathbb{N}\cup{0}.
\end{equation}
The decomposition (\ref{udek}) is straightforward. 

\subsection{ $A_2=\mathfrak{sl}(3,\mathbb{C})$ }   

The decomposition of the enveloping algebra of $A_2$ has already been considered by various authors, using both analytical and algebraic tools (\cite{Fla,Bur1} and references therein). Here we proceed somewhat differently, looking for the algebraic structure of commutants.  

\medskip Starting from the defining representation $[1,0]$, the Lie algebra $\mathfrak{sl}(3,\mathbb{C})$ is best given in terms of the elementary matrices $E_{ij}$ defined by  
\begin{equation*}
\left(E_{ij}\right)_{kl}=\delta_{i}^{k}\delta_{j}^{l},1\leq i,j,k,l\leq 3.
\end{equation*}
and subjected to the constraint ${\rm Tr}(X)=0$ for each $X\in \mathfrak{sl}(3,\mathbb{C})$. Table 1 gives the commutation relations in terms of these generators.  
 \begin{table}[htp]\label{TX1}
\caption{Commutators of the elementary matrices}
\begin{center}
\begin{tabular}{c|ccccccccc}
$\left[\circ,\circ\right]$ & $E_{11}$ & $E_{22}$ & $E_{33}$ & $E_{12}$ & $E_{13}$ & $E_{23}$ & $E_{21}$ & $E_{31}$ & $E_{32}$\\\hline
$E_{11}$& 0 & 0 & 0& $E_{12}$ & $E_{13}$ & $0$ & $-E_{21}$ & $-E_{31}$ & $0$\\
$E_{22}$& 0 & 0 & 0& $-E_{12}$ & $0$ & $E_{23}$ & $E_{21}$ & $0$ & $-E_{32}$\\
$E_{33}$& 0 & 0 & 0& $0$ & $-E_{13}$ & $-E_{23}$ & $0$ & $E_{31}$ & $E_{32}$\\
$E_{12}$& $-E_{12}$ & $E_{12}$ & 0& $0$ & $0$ & $E_{13}$ & $E_{11}-E_{22}$ & $-E_{32}$ & $0$\\
$E_{13}$& $-E_{13}$ & $0$ & $E_{13}$ & $0$ & $0$ & $0$ & $-E_{23}$ & $E_{11}-E_{33}$ & $E_{12}$\\
$E_{23}$& $0$ & $-E_{23}$ & $E_{23}$ & $-E_{13}$ & $0$ & $0$ & $0$ & $E_{21}$ & $E_{22}-E_{33}$\\
$E_{21}$& $E_{21}$ & $-E_{21}$ & $0$ & $E_{22}-E_{11}$ & $E_{23}$ & $0$ & $0$ & $0$ & $-E_{31}$\\
$E_{31}$& $E_{31}$ & $0$ & $-E_{31}$ & $E_{32}$ & $E_{33}-E_{11}$ & $-E_{21}$ & $0$ & $0$ & $0$\\
$E_{32}$& $0$ & $E_{32}$ & $-E_{31}$ & $0$ & $-E_{12}$ & $E_{33}-E_{22}$ & $E_{31}$ & $0$ & $0$\\
\end{tabular}
\end{center}
\label{TA2}
\end{table}%

Without loss of generality, we consider the Cartan subalgebra $\mathfrak{h}$ generated by $H_1=E_{11}-E_{22}$ and $H_2=E_{22}-E_{33}$. We observe that, due to this choice, $E_{12}$ and $E_{23}$ are the generators associated to the simple roots $\alpha_1,\alpha_2$ of $\Delta$, while $E_{13}$ is associated to the highest weight $\alpha_1+\alpha_2$ of the adjoint representation. Similarly, $E_{21},E_{32}$ and $E_{31}$ correspond to the negative roots $-\alpha_1$, $-\alpha_2$ and $-(\alpha_1+\alpha_2)$, respectively. Now   $E_{12}$ and $E_{23}$ generate a three-dimensional nilpotent algebra $\mathfrak{n}$.\footnote{It is indeed isomorphic to the Heisenberg algebra $\mathfrak{H}_1$.}, the centre $E_{13}$ of which, as an invariant, commutes with all generators of $\mathfrak{n}$. There are two other types of elements in the commutant $\mathcal{C}_{\mathcal{U}(\mathfrak{s})}(\mathfrak{n})$, namely the (generalized Casimir) invariants of $\mathfrak{s}$, that commute with all generators of the Lie algebra, and those polynomials that commute with the subalgebra $\mathfrak{n}$, but not with all generators of $\mathfrak{s}$, corresponding to subgroup scalars for the coadjoint representation. Using the analytical method (see \cite{Pe,C97}) it can be easily shown that an integrity basis of functions that commute with $\mathfrak{n}$ is formed by five independent elements. A direct computation of the polynomials of order at most three in the generators of $\mathfrak{s}$ shows that there are exactly six linearly independent elements, given respectively by 
\begin{equation}\label{fa1}
\begin{split}
B_1=& E_{13},\quad B_2= 3 E_{12} E_{23} + (H_1-H_2)E_{13},\\
C_1=&  E_{32} E_{13}^2 - E_{23} E_{12}^2 + H_2 E_{12} E_{13},\quad C_2= - E_{21} E_{13}^2 + E_{12} E_{23}^2 + H_1 E_{23} E_{13},\\
I_2=&  \frac{1}{3} (H_1^2 + H_1 H_2 + H_2^2) + H_1 + H_2 + E_{21} E_{12} +
E_{31} E_{13} + E_{32} E_{23},\\
I_{3} = & \frac{2}{27} H_1^3 + \frac{1}{9} H_1^2 H_2 - \frac{1}{9} H_1 H_2^2 - \frac{2}{27} H_2^3 + E_{32} E_{21} E_{13} + E_{31} E_{12} E_{23}- E_{31} E_{13}- E_{32} E_{23} \\
& + \frac{1}{3} E_{31} (H_1-H_2)E_{13} + \frac{1}{3} E_{21} (H_1 +2 H_2) E_{12} - \frac{1}{3} E_{32} ( 2H_1 + H_2) E_{23} + E_{21} E_{12} \\
& + \frac{1}{3} (H_1^2 - H_2^2) + \frac{1}{3} (H_1 - H_2).
\end{split}
\end{equation}
We observe that, with respect to the Cartan subalgebra $\mathfrak{h}$, the polynomials $B_1$ and $B_2$ have weight (1,1) , while $C_1$ is of weight (3,0) and $C_2$ of weight (0,3) respectively. The Casimir operators $I_2$ and $I_3$ are obviously of weight (0,0). As observed, at most five among these six operators are functionally independent, as follows from the algebraic dependence relation  
\begin{equation}\label{cons1}
C_1 C_2 + B_1^3 ( I_2 + I_3 ) - \frac{1}{3} B_1^2 B_2 ( I_2 -2 ) - \frac{1}{3} B_1 B_2^2 + \frac{1}{27} B_2^3 =0.
\end{equation}
The set (\ref{fa1}) contains an integrity basis. As Casimir operators of $\mathfrak{s}$, $I_2$ and $I_3$ generate an Abelian ideal, to which  $B_1,B_2$ can be added, thus exhausting the maximal number of commuting elements within the commutant (see equation (\ref{abel})). Therefore, the pair $C_1,C_2$ must lead to nonvanishing commutators. We obtain that $B_1,B_2,C_1,C_2$ generate a five-dimensional quartic polynomial algebra with nontrivial commutators 
\begin{equation}
\begin{split}
 [B_2,C_1]= 3 B_1 C_1,\quad [B_2,C_2]= -3 B_1 C_2,,\\
 [C_1,C_2]= - B_1^3 I_2 - B_1^2 B_2 + \frac{1}{3} B_1 B_2^2 
\end{split}
\end{equation}
It can be verified (see \cite{Fla}) that any other polynomial of order $d\geq 4$ is obtained as an algebraic expression in terms of $B_1,B_2,C_1,C_2,I_2,I_3$, 
showing that any element of the commutant $\mathcal{C}_{\mathcal{U}(\mathfrak{s})}(\mathfrak{n})$ can be written as 
\begin{equation}
P= B_1^{a_1}B_2^{a_2}C_1^{a_3}C_2^{a_4}I_2^{a_5}I_3^{a_6},\quad a_i\in\mathbb{N}\cup{0}
\end{equation}
subjected to the constraint $a_3a_4=0$, as a consequence of equation (\ref{cons1}). We also observe that the five elements $B_1,B_2,C_1,C_2,I_2$ are functionally independent and thus can be chosen as an integrity basis. As follows from the commutators, they also form a polynomial subalgebra $\mathcal{A}$ of the commutant.

\subsection{ $B_2=\mathfrak{so}(5,\mathbb{C})$ }

For the orthogonal Lie algebra $\mathfrak{so}(5,\mathbb{C})$, the root system $R$ has rank two with positive roots 
\begin{equation}
R_{+}=\left\{\alpha_1,\alpha_2,\alpha_1+\alpha_2,2\alpha_1+\alpha_2\right\} 
\end{equation}
Identifying the generators of the positive roots spaces with the elements $E_1,E_2,E_3,E_4$ and the negative roots with $F_1,F_2,F_3,F_4$, as well as the generators $H_1,H_2$ of the Cartan subalgebra $\mathfrak{h}$, the commutators are given in the following table:  

\begin{table}[htp]
\caption{Commutators of  $\mathfrak{so}(5,\mathbb{C})$}
\begin{center}
\begin{tabular}{c|cccccccccc}
$\left[\circ,\circ\right]$ & $H_{1}$ & $H_{2}$ & $E_{1}$ & $E_{2}$ & $E_{3}$ & $E_{4}$ & $F_{1}$ & $F_{2}$ & $F_{3}$ & $F_{4}$\\\hline
$H_{1}$& 0 & 0 & 0& $E_{2}$ & $E_{3}$ & $E_4$ & $0$ & $-F_{2}$ & $-F_3$ & $-F_{4}$\\
$H_{2}$& 0 & 0 & $E_1$ & $-E_{2}$ & $0$ & $E_4$ & $-F_1$ & $F_{2}$ & $0$ & $-F_{4}$\\
$E_{1}$& 0 & $0$ & $0$ & $E_{3}$ & $E_4$ & $0$ & $H_2$ & $0$ & $-F_2$ & $-F_{3}$\\
$E_{2}$& $-E_2$ & $E_2$ & $-E_3$ & $0$ & $0$ & $0$ & $0$ & $H_1-H_2$ & $F_1$ & $0$\\
$E_{3}$& $-E_3$ & $0$ & $-E_4$ & $0$ & $0$ & $0$ & $-E_2$ & $E_1$ & $H_1$ & $F_1$\\
$E_{4}$& $-E_4$ & $-E_4$ & $0$ & $0$ & $0$ & $0$ & $-E_3$ & $0$ & $E_1$ & $H_1+H_2$\\
$F_{1}$& $0$ & $F_1$ & $-H_2$ & $0$ & $E_2$ & $E_3$ & $0$ & $-F_3$ & $-F_4$ & $0$\\
$F_{2}$& $F_2$ & $-F_2$ & $0$ & $H_2-H_1$ & $-E_1$ & $0$ & $F_3$ & $0$ & $0$ & $0$\\
$F_{3}$& $F_3$ & $0$ & $F_2$ & $-F_1$ & $-H_1$ & $-E_1$ & $F_4$ & $0$ & $0$ & $0$\\
$F_{4}$& $F_4$ & $F_4$ & $F_3$ & $0$ & $-F_1$ & $-H_1-H_2$ & $0$ & $0$ & $0$ & $0$\\
\end{tabular}
\end{center}
\label{TA3}
\end{table}%
The nilradical $\mathfrak{n}$ of the Borel subalgebra, generated by $E_1,E_2,E_3,E_4$, is isomorphic to the only indecomposable nilpotent Lie algebra of dimension four $\mathfrak{n}_{4,1}$, and possesses two Casimir operators that are linear and quadratic in the generators, given by $P_1=E_4$ and $P_2=E_2E_4-\frac{1}{2}E_3^3$ respectively (see \cite{Sn14}). The weight of these polynomials are (1,1) and (2,0) respectively. Using formula (\ref{Rep2}), it can be easily verified that there are six algebraically independent solutions to the system. In addition to the two previous solutions $P_1$ and $P_2$, up to order four there are three additional polynomials that commute with the subalgebra $\mathfrak{n}$ but not with the whole orthogonal algebra:  
\begin{equation}
\begin{split}
P_3= &E_3-E_1E_2+E_4F_1-H_2E_3,\\
P_4 = & -E_4 H_1 + E_4 H_2 - 2 E_2 E_1^2 + 2 E_3 E_1 H_1 -  2 E_3 E_1 H_2 - E_4 H_1^2 + 2 E_4 H_2 H_1 \\
& - E_4 H_2^2 + 2 F_2 E_3^2 - 4 F_2 E_4 E_2,\\
P_5= & -2 E_4 E_3 - \frac{1}{2} E_3^2 E_1 + \frac{5}{2}  E_4 E_2 E_1 - E_4 E_3 H_1 + F_1 E_4^2 + \frac{1}{2} E_3 E_2 E_1^2-  \frac{1}{2} E_3^2 E_1 H_1 \\
& + \frac{1}{2} E_3^2 E_1  H_2 + \frac{1}{2} E_4 E_2 E_1 H_1 + \frac{1}{2} E_4 E_2 E_1 H_2- \frac{1}{2} E_4 E_3 H_2 H_1 + \frac{1}{2} E_4 E_3 H_2^2 - \frac{1}{2} F_1 E_4 E_3 E_1\\
& + \frac{1}{2} F_1 E_4^2 H_1  - \frac{1}{2} F_1 E_4^2 H_2 - \frac{1}{2} F_2 E_3^3 + 
  F_2 E_4 E_3 E_2 - \frac{1}{2} F_3 E_4 E_3^2 + 
  F_3 E_4^2 E_2
\end{split}
\end{equation}
Finally, the two Casimir operators $C_2$ and $C_4$ are given by 
\begin{equation}
\begin{split}
C_2 =&  H_1 + \frac{1}{3} H_2 - \frac{2}{3} E_1 F_1 - \frac{2}{3} E_2 F_2 - \frac{2}{3} E_3 F_3 - \frac{2}{3} 
 E_4 F_4 -\frac{1}{3} H_1^2 - \frac{1}{3} H_2^2,\\
 C_4= &  -H_2 - 2 F_1 E_1 - 2 F_4 E_4 -\frac{3}{2} H_2 H_1 - H_2^2 - 3 F_1 E_1 H_1 + 2 F_2 E_2 H_2 - 3 F_2 F_1  E_3\\
 & - 2 F_3 E_2 E_1 - F_3 E_3 H_2 - 2 F_3 F_1 E_4 -  F_4  E_4 H_1 - 2 F_4 E_4 H_2 - \frac{1}{2} H_2 H_1^2 -\frac{3}{2}  H_1H_2^2\\
 & - F_1 E_1 H_1^2 + F_2 E_2 H_2 H_1 - F_2 F_1 E_2 E_1 - F_2 F_1 E_3 H_1 - F_2 F_1 E_3 H_2 +  F_2 F_1^2 E_4 -\frac{1}{2} F_2^2 E_2^2\\
 & - F_3 E_2 E_1 H_1- F_3 E_2 E_1 H_2 - F_3 E_3 H_2^2 - 
 F_3 F_1 E_4 H_1 + F_3 F_1 E_4 H_2 - F_3 F_2 E_3 E_2 - 
 F_3^2 E_4 E_2\\
 & + F_4 E_2 E_1^2 - F_4 E_3 E_1 H_1 + F_4 E_3 E_1 H_2 - F_4 E_4 H_2 H_1 - F_4 F_1 E_4 E_1 -  F_4 F_2 E_3^2 + F_4  F_2 E_4 E_2\\
 & - F_4 F_3 E_4 E_3 - \frac{1}{2} F_4^2 E_4^2 - \frac{1}{2} H_1^2 H_2^2.
 \end{split}
\end{equation}
These operators exhaust the number of linearly independent elements in the commutant, and thus generate $\mathcal{C}_{\mathcal{U}(\mathfrak{s})}(\mathfrak{n})$ (see also \cite{Bur2} for a derivation using the Poisson bracket formalism). Among these seven operators, only six are algebraically independent, as follows from the relation (see \cite{Bur2}) 
\begin{equation}\label{cons2}
P_5^2-P_2P_4+2P_1P_3^2+P_3^3P_4-4P_1P_2+P_1P_2(6-P_4)-P_2P_4+P_1^2P_2C_4=0.
\end{equation}
The only nonvanishing commutators among these operators are given by 
\begin{equation}
\begin{split}
[P_3,P_4]=&- 2 P_1 P_3 + 4 P_5,\quad  [P_4,P_5]= 2 P_1^2 P_3 + P_1 P_3 P_4 - 4 P_1 P_5,\\
[P_3,P_5]= & - P_1 P_2 + \frac{3}{2} P_1 P_2C_2 - \frac{1}{2} P_1 P_3^2 - \frac{1}{2} P_2 P_4.  
\end{split}
\end{equation}
It follows that any element $P$ in the commutant can be written as 
\begin{equation}
P= P_1^{a_1}P_2^{a_2}P_3^{a_3}P_4^{a_4}P_5^{a_5}C_2^{a_6}C_4^{a_7},\quad a_i\in\mathbb{N}\cup{0}
\end{equation}
subjected to the constraint $a_5=0,1$ due to relation (\ref{cons2}). 

\medskip

We observe that, according to formula (\ref{abel}), at most five of the commutant generators commute with each other. In particular, the polynomials $\left\{P_1,\dots ,P_5,C_2\right\}$, which can be taken as an integrity basis for the system (\ref{Rep2}), further generate a polynomial (cubic) subalgebra of $\mathcal{C}_{\mathcal{U}(\mathfrak{s})}(\mathfrak{n})$. 

\section{The exceptional Lie algebra $G_2$}
 
The last of the (reduced) root systems of rank two corresponds to the $14$-dimensional exceptional Lie algebra $G_2$, with the set of positive roots given by
\begin{equation}
R_{+}=\left\{\alpha_1,\alpha_2,\alpha_1+\alpha_2,2\alpha_1+\alpha_2,3\alpha_1+\alpha_2,3\alpha_1+2\alpha_2\right\} 
\end{equation}   
Let $X_1,X_2$ denote the generators of the Cartan subalgebra and $X_3,X_5,X_7,X_9,X_{11},X_{13}$ be the generators associated to the positive roots, with  $X_4X_6,X_8,X_{10},X_{12},X_{14}$ being those corresponding to the negative roots. Over this basis, the commutators are given by 

\begin{table}[htp]
{\footnotesize
\caption{Commutators of  $G_2$}
\begin{center}
\begin{tabular}{c|cccccccccccc}
$\left[ ,\right] $ &  $X_{3}$ & $X_{4}$ & $X_{5}$ & $X_{6}
$ & $X_{7}$ & $X_{8}$ & $X_{9}$ & $X_{10}$ & $X_{11}$ & $X_{12}$ & $X_{13}$
& $X_{14}$ \\ 
$X_{1}$ &  $2X_{3}$ & $-2X_{4}$ & $-3X_{5}$ & $3X_{6}$ & $-X_{7}$
& $X_{8}$ & $X_{9}$ & $-X_{10}$ & $3X_{11}$ & $-3X_{12}$ & $0$ & $0$ \\ 
$X_{2}$ &  $-X_{3}$ & $X_{4}$ & $2X_{5}$ & $-2X_{6}$ & $X_{7}$ & $%
-X_{8}$ & $0$ & $0$ & $-X_{11}$ & $X_{12}$ & $X_{13}$ & $-X_{14}$ \\ 
$X_{3}$ &  $0$ & $X_{1}$ & $X_{7}$ & $0$ & $2X_{9}$ & $-3X_{6}$ & $%
-3X_{11}$ & $-2X_{8}$ & $0$ & $X_{10}$ & $0$ & $0$ \\ 
$X_{4}$   &  & $0$ & $0$ & $-X_{8}$ & $3X_{5}$ & $-2X_{10}$ & $2X_{7}$ & 
$3X_{12}$ & $-X_{9}$ & $0$ & $0$ & $0$ \\ 
$X_{5}$   &  &  & $0$ & $X_{2}$ & $0$ & $X_{4}$ & $0$ & $0$ & $-X_{13}$
& $0$ & $0$ & $X_{12}$ \\ 
$X_{6}$   &  &  &  & $0$ & $-X_{3}$ & $0$ & $0$ & $0$ & $0$ & $X_{14}$ & 
$-X_{11}$ & $0$ \\ 
$X_{7}$   &  &  &  &  & $0$ & $X_{1}+3X_{2}$ & $-3X_{13}$ & $2X_{4}$ & $0
$ & $0$ & $0$ & $X_{10}$ \\ 
$X_{8}$   &  &  &  &  &  & $0$ & $-2X_{3}$ & $3X_{14}$ & $0$ & $0$ & $%
-X_{9}$ & $0$ \\ 
$X_{9}$   &  &  &  &  &  &  & $0$ & $2X_{1}+3X_{2}$ & $0$ & $-X_{4}$ & $0
$ & $-X_{8}$ \\ 
$X_{10}$   &  &  &  &  &  &  &  & $0$ & $X_{3}$ & $0$ & $X_{7}$ & $0$ \\ 
$X_{11}$   &  &  &  &  &  &  &  &  & $0$ & $X_{1}+X_{2}$ & $0$ & $-X_{6}$
\\ 
$X_{12}$   &  &  &  &  &  &  &  &  &  & $0$ & $X_{5}$ & $0$ \\ 
$X_{13}$   &  &  &  &  &  &  &  &  &  &  & $0$ & $X_{1}+2X_{2}$ \\ 
$X_{14}$   &  &  &  &  &  &  &  &  &  &  &  & $0$\end{tabular}
\end{center}
\label{TA4} }
\end{table}%

The nilradical $\mathfrak{n}$ of the Borel subalgebra of $G_2$, generated by $X_3,X_5$, is of dimension six and isomorphic to the nilpotent Lie algebra $\mathfrak{n}_{6,19}$ listed in \cite{Sn14}. It possesses two Casimir operators of degrees one and two given by $Q_1=X_{13}$ and $Q_2=X_9^2-3X_3X_{13}+3X_7X_{11}$, respectively, and their weight with respect to the Cartan subalgebra $\mathfrak{h}=\mathbb{R}\langle X_1,X_2\rangle$ is given by (0,1) and (2,0). 
The Casimir operators $C_2=5 X_1 + 15 X_2 +  X_1^2 +  X_{10} X_9 + 3X_{12} X_{11} + 
 3 X_{14} X_{13} +3 X_2 X_1 + 3X_2^2 +  X_4 X_3 + 3X_6 X_5 +  X_8  X_7$ and $C_6$ of $G_2$ have degrees $2$ and $6$ (the explicit expression of the latter operator is skipped because of its length) and both have weight (0,0). These four polynomials commute with each other. As an integrity basis for the system (\ref{Rep2}) is formed by eight polynomials, four additional independent polynomials must be found. From these eight operators, at most six can commute, as follows from equation (\ref{abel}). We will see that, in this case, the commutant $\mathcal{C}_{\mathcal{U}(\mathfrak{s})}(\mathfrak{n})$ requires more than twice this number of operators, making the decomposition of the enveloping algebra of $G_2$ a computationally demanding problem. 

\medskip
For computational simplicity, and due to the dimension and the length of the polynomials commuting with $X_3,X_5$, it is convenient to proceed searching for polynomials with a given degree $d$ and weight $(\lambda,\mu)$, that will be denoted by $O_d^{[\lambda,\mu]}$. The strategy is to find an integrity basis formed by polynomials of lowest possible order, so that it contains a set of six independent commuting polynomials, and to complete it to a set of linearly independent polynomials.  

Up to degree four, the following seven polynomials in the commutant are linearly independent: 

\begin{itemize}

\item $d=1$: $O_1^{[0,1]}=Q_1$  

\item $d=2$: $O_2^{[2,0]}=Q_2$, $O_2^{[0,0]}=C_2$, 

\item $d=3$: 
\begin{equation*}
\begin{split}
O_3^{[3,0]}= & 2X_9^3 +27  X_{11}^2 X_5  - 27 
   X_{13}  X_{11}  X_2 -  27  X_{13}^2 X_6 +  9 X_{11} X_9  X_7- 9 X_{13} X_9 X_3,\\
   & \\
O_3^{[1,0]}= &    X_{10} X_9^2 + 3 (X_{11} X_{10} X_7 +  X_{11} X_4 X_1-  X_{13} X_{10} X_3) + 
  9 (X_{11} X_4 X_2+  X_{11} X_8 X_5+ X_{13} X_6 X_4)\\
  & + 3 (X_{13} X_8 X_1+X_7 X_3 X_2 - X_5 X_3^2+ X_7^2 X_6+X_9 X_2 X_1) + X_9 X_4 X_3  +   X_9 X_8 X_7    + X_9 X_1^2\\
  & - 9 X_9 X_6 X_5 + 2 X_9 + 3 X_11 X_4 - 6 X_{13} X_8 - 6 X_7 X_3 + X_9 X_1.  
\end{split}
\end{equation*}

\item $d=4$: 
\begin{equation*}
\begin{split}
O_4^{[2,0]}= &    2 X_9^2X_{13}X_{14}+2 X_ {9}^2 X_ {11} X_ {12}-3 X_ {8}^2 X_ {13} X_ {13}-2 X_ {7}^2 X_ {8} X_ {11} - X_ {7} X_ {9} X_ {10} X_ {11}+6 X_ {7} X_ {11} X_ {13} X_ {14}\\ 
&  +6 X_ {5} X_ {8} X_ {9} X_ {11} + 9 X_ {6} X_ {10} X_ {13}^2 +
 X_ {6} X_ {7}^2 X_ {9} + 6 X_ {7} X_ {11}^2X_ {12} -4 X_ {5} X_ {6} X_ {9}^2-3 X_ {5} X_ {6} X_ {7} X_ {11} - 3 X_ {4}^2 X_ {11} ^2\\
  &+X_ {3} X_ {9} X_ {10} X_ {13}  + 
 6 X_ {4} X_ {6} X_ {9} X_ {13} - 6 X_ {4} X_ {8} X_ {11} X_ {13} - 
 9 X_ {5} X_ {10} X_ {11} X_ {11}+2 X_ {3} X_ {7} X_ {8} X_ {13} -X_ {2}^2 X_ {9}^2  \\
 &+2 X_ {3}^2 X_ {4} X_ {13} - X_ {3}^2 X_ {5} X_ {9} - 
 2 X_ {3} X_ {4} X_ {7} X_ {11} + 3 X_ {3} X_ {5} X_ {6} X_ {13} -6 X_ {3} X_ {11} X_ {12} X_ {13} - 6 X_ {3} X_ {13}^2 X_ {14} \\
   & +9 X_ {2} X_ {3} X_ {5} X_ {11} + X_ {2} X_ {3} X_ {7} X_ {9} - 
 9 X_ {2} X_ {6} X_ {7} X_ {13} - 6 X_ {2} X_ {8} X_ {9} X_ {13} -3 X_ {2}^{2} X_ {7} X_ {11}-2 X_ {1} X_ {8} X_ {9} X_ {13} \\
& -X_ {1}^2X_ {3} X_ {13} + X_ {1}^2 X_ {7} X_ {11} - 
 6 X_ {1} X_ {2} X_ {3} X_ {13} +6 X_ {1} X_ {3} X_ {5} X_ {11} - 2 X_ {1} X_ {4} X_ {9} X_ {11} -6 X_ {1} X_ {6} X_ {7} X_ {13} \\
& +\left(9 X_ {2} X_ {10} X_ {11}  - 6 X_ {2}^2 X_ {3}  +5 X_ {1} X_ {3}\right) X_ {13} - 11 X_ {1} X_ {7} X_ {11}-6 X_ {10} X_ {11} X_ {13}+6 X_ {2} X_ {3} X_ {13}-4 X_ {1} X_ {9}^2\\
    &+\left(12 X_ {3} X_ {5}   -15 X_ {2} X_ {7}\right) X_ {11}+2 X_ {3} X_ {7} X_ {9}-2 X_ {4} X_ {9} X_ {11}-3 X_ {6} X_ {7} X_ {13}  -5 X_ {2} X_ {9}^2 
    -5 X_ {7} X_ {11} + \frac{4}{3}X_ {9}^2.           \\
     & \\
  O_4^{[0,2]}= & X_ {7}^2 X_ {9}^2+4 X_ {7}^3 X_ {11} -18 X_ {5} X_ {7} X_ {9} X_ {11}-18 X_ {1} X_ {5} X_ {11} X_ {13}-2 X_ {1} X_ {7} X_ {9} X_ {13} - 12 X_ {3} X_ {4} X_ {13}^2-4 X_ {5} X_ {9}^3 \\
     & + X_ {3} \left(12X_ {5} X_ {9}  - 4  X_ {7}^2\right)X_ {13}+12 X_ {4} X_ {7} X_ {11} X_ {13} + 4 X_ {4} X_ {9}^2 X_ {13}-3 X_ {1}^2X_ {13}^2    - 
 27 X_ {5}^2 X_ {11}^2-6 X_ {1} X_ {13}^2 \\
     & -81 X_ {5} X_ {11} X_ {13} - 3 X_ {7} X_ {9} X_ {13}.  \\
  \end{split}
\end{equation*}

\end{itemize}

These polynomials, jointly with the Casimir operator $O_6^{[0,0]}=C_6$ of $G_2$, are algebraically independent and can thus be considered as an integrity basis $\mathcal{I}$ of the system (\ref{Rep2}) associated to the nilpotent Lie algebra $\mathfrak{n}$. Starting from this set $\mathcal{I}$, we analyze the commutators in order to obtain a basis for the commutant. To this extent, we relabel $C_2$ as $Q_3$ and the cubic and quartic polynomials as 
\begin{equation*}
Q_4=O_3^{[3,0]},\quad Q_5=O_3^{[1,0]},\quad Q_6=O_4^{[2,0]},\quad Q_7=O_4^{[0,2]}.
\end{equation*}
 It is immediate to verify that $Q_1,\dots Q_5$ form an Abelian algebra. Adding the sixth-order Casimir operator of $G_2$, we obtain the maximal number of operators in $\mathcal{C}_{\mathcal{U}(\mathfrak{s})}(\mathfrak{n})$ that commute with each other (see equation (\ref{abel})). Considering now the commutator of $Q_4$ and $Q_7$, we observe that it decomposes as 
 \begin{equation}\label{kla1}
\left[Q_4,Q_7\right]= -54Q_1 O_5^{[3,1]}, 
\end{equation}
where 
\begin{equation*}
\begin{split}
O_5^{[3,1]}= &    2 X_8X_9^2 X_{13}^2+ 
 6 X_7X_8 X_{11}X_{13}^2+ X_7^2 X_9^2 X_{11}+ 
 4 X_{7}^3 X_{11}^2   + 3 X_{6}  X_{7}  X_{9}  X_{13}^2- 
 4 X_{5}  X_{9}^3  X_{11} - 18 X_{5}  X_{7}  X_{9}  X_{11} ^2\\
 & + 
 27 X_{5}  X_{6}  X_{11}  X_{13}^2- 27 X_{5}^2  X_{11}^3+ 
 2 X_{4}  X_{9}^2 X_{11}  X_{13}  + 6 X_{4}  X_{7}  X_{11}^2 X_{13} - 
 6 X_{3}  X_{8}  X_{13}^3- X_{3}  X_{7}  X_{9}^2  X_{13} \\
 &   - 
 8 X_{3}  X_{7}^2  X_{11}  X_{13} + 15 X_{3}  X_{5}  X_{9}  X_{11}  X_{13}  - 
 6 X_{3}  X_{4}  X_{11}  X_{13}^2+ 4 X_{3}^2  X_{7}  X_{13}^2+ 
  X_{2} (2 X_{9}^3   + 9  X_{7}  X_{9}  X_{11})  X_{13}\\
 &  + 
 27 X_{2}  X_{5}  X_{11}^2  X_{13}  - 6 X_{2}  X_{3}  X_{9}  X_{13}^2- 
 X_{1}  X_{7}  X_{9}  X_{11}  X_{13}  + (9 X_{1}  X_{6}  X_{13}^2 - 
 9 X_{1}  X_{5}  X_{11}^2)  X_{13} + \frac{2}{3} X_{9}^3  X_{13} \\
 & + (X_{1}  X_{3}  X_{9}  X_{13}+ 
 9 X_{1}  X_{2}  X_{11}  X_{13} - 7 X_{3}  X_{9}  X_{13}+24 X_{11}  X_{13}- 
 72 X_{5}  X_{11}^2)X_{13}.
\end{split}
\end{equation*}
This fifth-order polynomial is linearly independent of  $\mathcal{I}$, hence must be added to the set. Two further fifth-order operators are obtained by successive commutators:\footnote{For these as well as the following operators, we omit their  explicit expression due to their length.}  
\begin{equation}\label{kla2}
\left[Q_5,Q_7\right]= 27Q_1 O_5^{[1,1]},\quad  \left[Q_5,O_5^{[1,1]}\right]= Q_1 Q_2Q_3^2-Q_1Q_5^2-6Q_2 O_5^{[0,1]}, 
\end{equation}
The operators in $\mathcal{J}_1=\mathcal{I}\cup \left\{O_5^{[0,1]},O_5^{[1,1]},O_5^{[3,1]}\right\}$ are linearly independent, and exhaust the polynomials of order $d\leq 5$ having this property. We label them as $Q_8=O_5^{[0,1]},\; Q_9=O_5^{[1,1]},\; Q_{10}=O_5^{[3,1]}$. The commutator $\left[Q_5,Q_6\right]= 27 O_6^{[3,0]}$ provides a linearly independent operator of order six  that we label as $Q_{11}$. The commutator $\left[Q_6,Q_7\right]= 12 Q_1 O_6^{[2,1]}$ gives rise to an operator $Q_{12}=O_6^{[2,1]}$ that is also independent on the previous eleven polynomials. A routine computation shows that $Q_{11},Q_{12}$ and $Q_{13}=C_6$ are the only sixth-order polynomials linearly independent of the elements in $\mathcal{J}_1$. An operator of seventh-order not expressible in terms of the previous elements is obtained from the commutator
\begin{equation}\label{kla3}
\left[Q_5,Q_8\right]= 6O_7^{[1,1]}.
\end{equation}
We label this polynomial as $Q_{14}$. Two additional linearly independent operators are obtained from the commutators of the latter as 
\begin{equation}\label{kla4}
\left[Q_8,Q_{14}\right]= -\frac{4}{9}Q_3^2Q_5Q_7-\frac{1}{6}Q_1^2Q_5Q_{13}-\frac{4}{3}Q_5O_8^{[0,2]}-\frac{2}{9}Q_3O_9^{[1,2]} 
\end{equation}
We denote them as $Q_{15}=O_8^{[0,2]}$ and $Q_{16}=O_9^{[1,2]}$ respectively. Finally, the commutator $\left[Q_8,Q_{15}\right]=4O_{12}^{[0,3]}$ provides a polynomial of order twelve that is linearly independent of $\left\{Q_1,\dots Q_{16}\right\}$, and that we denote as $Q_{17}$. Analyzing higher orders does not lead to new polynomials that are linearly independent of the former. Any other commutator among these operators can hence be expressed in terms of $\left\{Q_1,\dots Q_{17}\right\}$. Excluding the commutators above, the remaining non-vanishing commutators are given by\footnote{Due to simplicity, we give the commutators and relations in its symmetric form.}
\begin{equation*}\label{kla5}
\begin{split}
\left[Q_4,Q_8\right]=& -27Q_1Q_{11},\qquad \left[Q_4,Q_9\right]=3Q_1\left(2Q_2Q_6-Q_4Q_5-6Q_2^2Q_3\right),\qquad \left[Q_4,Q_{10}\right]=Q_1\left(Q_4^2-4Q_2^3\right),\\
\left[Q_4,Q_{12}\right]=&3Q_1\left(Q_4Q_6-2Q_2^2Q_5-3Q_2Q_3Q_4\right),\qquad \left[Q_4,Q_{14}\right]=Q_1\left(Q_2Q_5^2-3Q_2^2Q_3^2+Q_3Q_4Q_5+4Q_2Q_3Q_6\right),\\
\left[Q_4,Q_{15}\right]=&3Q_1\left(Q_6Q_9-3Q_2Q_3Q_9+Q_5Q_{17}\right),\qquad \left[Q_4,Q_{16}\right]=18Q_1\left(Q_6Q_{12}-Q_2Q_3Q_9\right)-243Q_1^2Q_5Q_{11}\\
& -27Q_1\left(Q_3Q_4Q_9-3Q_3Q_5Q_{10}\right),\\
\left[Q_4,Q_{17}\right]=&\frac{9}{2}Q_1^3\left(2Q_3^2Q_5Q_6-Q_3^4Q_4+Q_2Q_5(Q_{13}-9Q_3^3)-Q_3Q_5^3\right)-\frac{3}{4}Q_1\left(Q_5^3Q_7+3Q_4Q_8^2-\frac{1}{2}Q_4Q_7Q_{13}\right)\\
& +\frac{27}{4}Q_1\left(Q_5Q_9^2-Q_2Q_3^2Q_5+Q_3Q_9Q_{12}+\frac{3}{2}Q_4Q_8^2\right)+\frac{27}{2}Q_1^2\left(Q_3^2Q_4Q_8-Q_5Q_6Q_8+\frac{9}{2}Q_2Q_3Q_5Q_8\right),\\ 
\left[Q_5,Q_{10}\right]=& \frac{1}{3}Q_1\left(Q_4Q_5-2Q_2Q_6+6Q_2^2Q_3\right),\quad \left[Q_5,Q_{11}\right]=\frac{2}{9}\left(Q_6^2-Q_3Q_4Q_5+2Q_2^2Q_3^2\right)-\frac{2}{9}Q_2\left(Q_5^2-4Q_3Q_6\right),\\
\left[Q_5,Q_{12}\right]=& 2Q_1\left(Q_3^2Q_4+3Q_2Q_3Q_5-Q_5Q_6\right)-3Q_4Q_8,\quad \\
\left[Q_5,Q_{14}\right]=&\frac{1}{3}Q_1Q_2\left(8Q_3^3-Q_{13}\right)+\frac{4}{3}Q_1Q_3\left(Q_5^2-Q_3Q_6\right) +2Q_6Q_8-4Q_2Q_3Q_8,\quad \left[Q_5,Q_{15}\right]=(2Q_1Q_3^2-3Q_8)Q_9,\quad  \\
\left[Q_5,Q_{16}\right]=& 12Q_1Q_3\left(Q_3Q_{12}-Q_5Q_9\right) +72Q_1Q_5Q_{14}-18Q_8Q_{12},\\
\left[Q_5,Q_{17}\right]=& -\frac{9}{2}Q_1^3Q_3^4Q_5+\frac{27}{2}Q_1^2Q_3^2Q_5Q_8-Q_1\left(Q_3^2Q_{16}+\frac{9}{2}Q_3Q_9Q_{15}-\frac{1}{8}Q_5Q_7Q_{13}+\frac{81}{8}Q_5Q_8^2+3Q_3^3Q_5Q_7\right)\\
&+\frac{9}{2}Q_3Q_5Q_7Q_8+\frac{3}{2}Q_8Q_{16},\qquad \left[Q_6,Q_{8}\right]=-2Q_5Q_9-2Q_3Q_{12},     \\
\left[Q_6,Q_{9}\right]=&2Q_1\left(Q_3^2Q_4+9Q_2Q_3Q_5-2Q_5Q_6\right)-3Q_4Q_8,\qquad \left[Q_6,Q_{10}\right]=\frac{2}{3}Q_1\left(Q_4Q_6-2Q_2^2Q_5-3Q_2Q_3Q_4\right),\\ 
\left[Q_6,Q_{11}\right]=&\frac{4}{9}Q_2\left(Q_5Q_6-4Q_2Q_3Q_5-\frac{2}{3}Q_3^2Q_4\right)+\frac{2}{9}Q_4\left(Q_3Q_6-Q_5^2\right),\quad \left[Q_6,Q_{12}\right]=2Q_1Q_6^2-6Q_2^2Q_8+2Q_1\times\\
 & \left(2Q_2^2Q_3^2  -3Q_2Q_5^2-3Q_2Q_3Q_6\right),\qquad \left[Q_6,Q_{15}\right]=2Q_1\left(Q_3^2Q_{12}+3Q_3Q_5Q_9+9Q_5Q_{14}\right)-3Q_8Q_{12},
 \\
 \left[Q_6,Q_{14}\right]=&\frac{5}{3}Q_1\left(Q_2Q_3^2Q_5+Q_5^3-Q_3Q_5Q_6-\frac{2}{5}Q_3^3Q_4+\frac{1}{10}Q_4Q_{13}\right) +Q_3Q_4Q_8+2Q_2Q_5Q_8, \\
 \left[Q_6,Q_{16}\right]=& 6Q_1\left(\left(2Q_2Q_3^2-Q_5^2+2Q_3Q_6\right)Q_9-9Q_3^3Q_{10}-81Q_8Q_{11}\right)+9Q_1\left(10Q_3Q_5Q_{12}-Q_{10}Q_{13}\right)+324Q_1^2\times\\
 & Q_3^2Q_{11}+81Q_3Q_8Q_{10}-18Q_2Q_8Q_9,\qquad  
\left[Q_6,Q_{17}\right]=3Q_1^3\left(Q_3^4Q_6-3Q_2Q_3^5+6Q_3^3Q_5^2-\frac{1}{2}Q_5^2Q_{13}\right)\\
 & +9Q_1^2\left(Q_3^2Q_6+3Q_2Q_3^3Q_8-3Q_3Q_5^2Q_8\right)-\frac{3}{4}Q_5^2Q_7Q_8+\frac{9}{4}Q_8Q_9^2+\frac{27}{4}Q_1\left(3Q_2Q_3Q_8^2-Q_6Q_8^2\right)\\
 & +5Q_1Q_3^2Q_5^2Q_7+3Q_1\left(Q_3^2Q_9^2-3Q_5^2Q_{15}-9Q_3Q_9Q_{14}\right)+\frac{1}{4}Q_1\left(Q_6Q_7Q_{13}-3Q_2Q_3Q_7Q_{13}\right),\\
 \left[Q_7,Q_{9}\right]=& 2Q_1Q_5Q_7,\quad  \left[Q_7,Q_{10}\right]=2Q_1\left(8Q_1^2Q_2Q_5-Q_4Q_7\right),\quad  \left[Q_7,Q_{11}\right]=-2Q_1Q_4Q_8+\frac{4}{3}Q_1^2\times\\
 & \left(Q_3^2Q_4+4Q_2Q_3Q_5-Q_5Q_6\right),\quad  \left[Q_7,Q_{12}\right]=12Q_1^2Q_2\left(2Q_1Q_3^2-3Q_8\right)+4Q_1\left(3Q_2Q_3-Q_6\right)Q_7,\\
 \left[Q_7,Q_{14}\right]=& -\frac{2}{3}Q_1\left(Q_{16}+4Q_3Q_5Q_7\right)-6Q_1^3Q_3^2Q_5,\quad \left[Q_7,Q_{16}\right]=3Q_1\left(8Q_3Q_7+18Q_1^2Q_3^2\right)Q_9+18Q_1Q_7Q_{14}\\
 & -81Q_1^2Q_8,\quad \left[Q_8,Q_{9}\right]=9Q_1Q_5Q_8-\frac{8}{3}Q_3Q_5Q_7-6Q_1^2Q_3^2Q_5-\frac{2}{3}Q_{16},\\
 \left[Q_8,Q_{9}\right]=& \frac{2}{3}Q_1^2\left(Q_3^2Q_4-Q_5Q_6+4Q_2Q_3Q_5\right)-Q_1Q_4Q_8,\quad \left[Q_8,Q_{11}\right]=\frac{8}{9}Q_1\left(Q_2Q_3^2+\frac{1}{3}Q_5^2\right)Q_5+\frac{1}{27}Q_1\times\\
 & Q_4Q_{13}-\frac{4}{9}Q_1Q_3Q_5Q_6,\quad 
\left[Q_8,Q_{12}\right]= 4Q_1^2Q_2Q_3^3-6Q_1Q_2Q_3Q_8-Q_9^2+(2Q_2Q_3^2+\frac{1}{3}Q_5^2-\frac{2}{3}Q_2Q_6)Q_7,\\
\left[Q_8,Q_{16}\right]=& 18Q_1^2Q_3^2\left(Q_3Q_9+3Q_{14}\right)+4Q_3^2Q_7Q_9-27Q_1\left(Q_3Q_8Q_9-3Q_8Q_{14}\right)+30Q_3Q_7Q_{14}-\frac{3}{2}Q_1^2Q_9Q_{13},\\
\end{split}
\end{equation*}
\begin{equation*}\label{kla5}
\begin{split}
 \left[Q_8,Q_{17}\right]=& 2Q_1^4Q_3^6-9Q_1^3Q_3^4Q_8+Q_1^2\left(Q_3^5Q_7+\frac{27}{2}Q_3^2Q_8^2+\frac{1}{6}Q_3^2Q_7-12Q_3^3Q_{15}+Q_{13}Q_{15}\right)-\frac{27}{4}Q_1Q_8^3\\
 & -3Q_1Q_3\left(Q_3^2Q_7-6Q_8Q_{15}\right)+\frac{1}{12}Q_3Q_7\left(Q_7Q_{13}+27Q_8^2-36Q_3Q_{15}\right)-\frac{1}{4}Q_1Q_4Q_7Q_{13}+6Q_{15}^2,\\
\left[Q_9,Q_{10}\right]=&-Q_1\left(Q_5Q_{10}+Q_4Q_9\right),\quad \left[Q_9,Q_{11}\right]=3Q_1Q_5Q_{11}+\frac{1}{3}Q_1\left(Q_4Q_9+5Q_5Q_{10}\right)+\frac{2}{9}\left(Q_2Q_5Q_9-Q_6Q_{12}\right),\\
\left[Q_9,Q_{12}\right]=& 3Q_1\left((3Q_2Q_3-Q_6)Q_9-2Q_3^2Q_{10}\right)+9Q_8Q_{10},\quad \left[Q_9,Q_{14}\right]= \frac{4}{3}Q_1Q_3\left(Q_5Q_9+Q_3Q_{12}\right)+Q_1Q_5Q_{14}\\
& -2Q_8Q_{12},\quad \left[Q_9,Q_{15}\right]=Q_5Q_7Q_8-\frac{2}{3}Q_1Q_3^2Q_5Q_7,\\
\left[Q_9,Q_{16}\right]=& 9Q_1^3\left(4Q_2Q_3^4-Q_3^2Q_5^2\right)+Q_1\left(108Q_2Q_3^2+\frac{27}{2}Q_5^2\right)Q_8+81Q_1\left(Q_2Q_8^2+Q_9Q_{14}\right)-6Q_6Q_7Q_8\\
& +Q_1\left(4Q_3^2Q_6Q_7+3Q_2Q_7Q_{13}\right)\\
\left[Q_9,Q_{17}\right]=& \frac{9}{2}Q_1^2Q_3^2\left(Q_1Q_3^2-3Q_8\right)Q_9+Q_1Q_3\left(Q_3^2Q_7Q_9+3Q_3Q_7Q_{14}-\frac{9}{2}Q_9Q_{15}\right)-\frac{3}{2}Q_3Q_7Q_8Q_9-\frac{9}{2}Q_7Q_8Q_{14}\\
& +\frac{1}{8}Q_1\left(Q_7Q_9Q_{13}+81Q_8^2Q_9\right),\quad \left[Q_{10},Q_{11}\right]=-Q_1Q_4Q_{11},\quad \left[Q_{10},Q_{12}\right]=Q_1\left(Q_6-3Q_2Q_3\right)Q_{10}\\
& +2Q_1Q_2^2Q_9,\quad \left[Q_{10},Q_{14}\right]=\frac{3}{2}Q_1^2Q_5Q_{11}-\frac{2}{9}Q_1\left(Q_6Q_{12}+2Q_2Q_5Q_9-3Q_3Q_5Q_{10}\right),\\
\left[Q_{10},Q_{15}\right]=& -\frac{4}{3}Q_1^3Q_2Q_3^2Q_5+2Q_1^2Q_2Q_5Q_8+Q_1\left(Q_9Q_{12}-Q_2Q_3Q_5Q_7+\frac{1}{3}Q_5Q_6Q_7\right), \\
\left[Q_{10},Q_{16}\right]=&Q_1^3\left(7Q_5^2Q_6-4Q_3^2Q_4Q_5+3Q_2^2Q_{13}+3Q_2Q_3\left(4Q_3Q_6+6Q_5^2-12Q_2Q_3^2\right)\right)+6Q_1^2\left(Q_4Q_5-3Q_2Q_6\right)Q_8\\
& +54Q_1^2Q_2^2Q_3Q_8+Q_1Q_2Q_3\left(8Q_6-9Q_2Q_3\right)Q_7-\frac{1}{3}Q_1\left(Q_2Q_5^2+5Q_6^2\right)Q_7+Q_1\left(6Q_2Q_9^2-5Q_3Q_4Q_5Q_7\right),\\
\left[Q_{10},Q_{17}\right]=& \frac{3}{2}Q_1^3Q_3\left(Q_5^2+5Q_2Q_3^2-\frac{2}{3}Q_3Q_6\right)Q_9+2Q_1^2Q_3^2\left(Q_5Q_{12}-\frac{9}{4}Q_3^2Q_{10}\right)-\frac{1}{2}Q_1^3Q_2Q_9Q_{13}-\frac{3}{4}Q_1Q_9^3\\
& +\frac{3}{2}Q_1^2\left(Q_6-\frac{15}{2}Q_2Q_3\right)Q_8Q_9-3Q_1^2\left(Q_5Q_8Q_{12}-\frac{9}{2}Q_3^2Q_8Q_{10}\right)+\frac{3}{4}Q_1Q_3\left(Q_3Q_5Q_7Q_{1}2-Q_6Q_7Q_9\right)\\
& +\frac{3}{4}Q_1\left(Q_5^2+3Q_2Q_3^2\right)Q_7Q_9-\frac{3}{8}Q_1\left(Q_7Q-{13}+27Q_8Q_{10}^2\right)\\
\left[Q_{11},Q_{12}\right]=& 3Q_1\left(Q_6-3Q_2Q_3\right)Q_{11}+\frac{2}{9}Q_2\left(Q_6-3Q_2Q_3\right)Q_9-\frac{1}{9}Q_4Q_5Q_9+\frac{1}{3}\left(Q_5^2+6Q_2Q_3^2-2Q_3Q_6\right)Q_{10},\\
\left[Q_{11},Q_{14}\right]=&\frac{1}{9}Q_3^2\left(Q_5Q_{10}-Q_4Q_9\right)+\frac{2}{27}\left(Q_3Q_6-Q_5^2\right)Q_{12}+\frac{2}{27}\left(Q_5Q_6-4Q_2Q_3Q_5\right)Q_9,\\
\left[Q_{11},Q_{15}\right]=&\frac{2}{9}Q_1^2\left(Q_3^4Q_4+13Q_2Q_3^3Q_5+6Q_3Q_5^3-4Q_3^2Q_5Q_6\right)-\frac{1}{3}Q_1^2Q_2Q_5Q_{13}-\frac{13}{3}Q_1Q_2Q_3Q_5Q_8-\frac{1}{3}Q_5Q_9^2\\
& +\frac{2}{3}Q_1\left(2Q_5Q_6-Q_3^2Q_4\right)Q_8-\frac{1}{9}Q_3\left(Q_5Q_6Q_7+3Q_9Q_{12}\right)+\frac{1}{54}\left(Q_4Q_{13}+2Q_5^3\right)Q_7+\frac{1}{2}Q_4Q_8^2\\
& +\frac{1}{3}Q_2Q_3^2Q_5Q_7, \\
\left[Q_{11},Q_{16}\right]=& \frac{1}{3}Q_1^2\left(68Q_2^2Q_3^4+8Q_3^3Q_4+26Q_2Q_3^2Q_5^2+2(Q_2Q_6-Q_4Q_5)Q_{13}+10Q_5^4-2Q_3(Q_5^2 +24Q_2Q_3^2)Q_6\right)\\
& +\frac{8}{3}Q_1^2Q_3^2Q_6^2-2Q_1^2Q_2^2Q_3Q_{13}-4Q_1\left(Q_6^2+Q_2Q_5^2+Q_3Q_4Q_5-6Q_2Q_3Q_6-8Q_2^2Q_3^2\right)Q_8+Q_6Q_9^2\\
& +3Q_2^2\left(2Q_3^3Q_7-Q_8^2-\frac{1}{27}Q_7Q_{13}\right)-\frac{1}{9}\left(Q_5^2Q_6-10Q_3Q_6^2+4Q_2Q_3Q_5^2-6Q_3^2Q_4Q_5+48Q_2Q_3^2Q_6  \right)Q_7\\
& -2Q-2Q_3Q_9^2,\\
\left[Q_{11},Q_{17}\right]=&\frac{1}{2}Q_1^2\left(Q_3^3Q_6-\frac{4}{3}Q_3^2Q_5^2-\frac{22}{3}Q_2Q_3^4+\frac{1}{2}Q_2Q_3Q_{13}-\frac{1}{6}Q_6Q_{13}\right)Q_9+\frac{27}{2}Q_1^2Q_3^2\left(Q_8Q_{11}-\frac{7}{81}Q_3Q_5Q_{12}\right)\\
& +Q_1Q_3^3\left(Q_1Q_3^2-3Q_8\right)Q_{10}+\frac{1}{6}\left(Q_3^2Q_6-\frac{4}{3}Q_3Q_5^2-3Q_2Q_3^3\right)Q_7Q_9 -\frac{1}{6}Q_3^2Q_5Q_7Q_{12}-\frac{9}{2}Q_1^3Q_3^4Q_{11}\\
&+\frac{1}{4}Q_1\left(7Q_2Q_3^2-3Q-3Q_6+8Q_5^2)\right)Q_8Q_9+\frac{3}{2}Q_2Q_8^2Q_9+\frac{9}{4}Q_3Q_8^2Q_{10}+\frac{7}{4}Q_1Q_3Q_5Q_8Q_{12}-\frac{81}{8}Q_1\times\\
& Q_8^2Q_{11}+\frac{1}{12}\left(Q_3Q_7+3Q_1^2Q_3^2\right)Q_{10}Q_{13}+\frac{1}{18}Q_2Q_7Q_9Q_{13}-\frac{3}{8}Q_1Q_8Q_{10}Q_{13}-Q_1^2Q_3Q_5^2Q_{14}-\frac{1}{6}\times\\
& \left(Q_5^2Q_7-9Q_9^2Q_{14}\right),\\
\end{split}
\end{equation*}
\begin{equation*}\label{kla5}
\begin{split}
\left[Q_{12},Q_{14}\right]=&\frac{9}{2}Q_1^2Q_3^2Q_{11}-\frac{1}{6}Q_1\left(Q_2Q_3^2+8Q_5^2-3Q_3Q_6\right)Q_9-2Q_2Q_8Q_9+\left(3Q_3Q_8-2Q_1Q_3^3\right)Q_{10}\\
& -\frac{27}{4}Q_1Q_8Q_{11}-\frac{1}{2}Q_1Q_3Q_5Q_{12}+\frac{1}{4}Q_1Q_{10}Q_{13},\\
\left[Q_{12},Q_{15}\right]=& -2Q_1^3Q_2Q_3^4+6Q_1^2Q_2Q_3^2Q_8+2Q_1\left(Q_2Q_3^3+\frac{3}{4}Q_3Q_5^2-\frac{1}{3}Q_3^2Q_6\right)Q_7+Q_6\left(Q_6-3Q_2Q_3\right)Q_7Q_8\\
& -\frac{1}{2}Q_1Q_2\left(Q_7Q_{13}+9Q_8^2\right)-\frac{3}{2}Q_1Q-3Q-9^2-18Q_1Q_9Q_{14},\\
\left[Q_{12},Q_{16}\right]=&3Q_1^3\left(7Q_3^4Q_4+25Q_2Q_3^3Q_5-4Q_3^2Q_5Q_6-3Q_2Q_5Q_{13}\right)+9Q_1^2\left(Q_5Q_6-7Q_3^2Q_4-\frac{45}{2}Q_2Q_3Q_5\right)Q_8\\
&+Q_1\left(6Q_3^3Q_4+\frac{89}{2}Q_2Q_3^2Q_5-\frac{9}{2}Q_5^3-\frac{27}{2}Q_3Q_5Q_6\right)Q_7-\left(9Q_3Q_4+6Q_2Q_5\right)Q_7Q_8-\frac{9}{2}Q_1Q_5Q_9^2\\
& +\frac{189}{4}Q_1Q_4Q_8^2-\frac{27}{2}Q_1Q_3Q_9Q_{12}+\frac{3}{4}Q_1Q_4Q_7Q_{13},\\
\left[Q_{12},Q_{17}\right]=&\frac{3}{2}Q_1^3Q_5Q_9Q_{13}-Q_1Q_3^2\left(Q_5Q_7+9Q_1^2Q_3Q_5\right)Q_9+9Q_1\left(Q_5Q_9+Q_3Q_{12}\right)Q_{15}+\frac{27}{2}Q_1^2Q_3Q_5Q_8Q_9\\
& -\frac{1}{4}Q_1\left(Q_7Q_{12}Q_{13}+27Q_8^2Q_{12}\right)-3Q_1^2Q_3^2\left(Q_1Q_3^2-3Q_8\right)Q_{12}+\frac{3}{2}Q_5Q_7Q_8Q_9,\\
\end{split}
\end{equation*}

\begin{equation*}\label{kla5}
\begin{split}
\left[Q_{14},Q_{15}\right]=&\frac{2}{3}\left(Q_8-\frac{2}{3}Q_1Q_3^2\right)Q_{16}-Q_1^3Q_3^4Q_5+3Q_1^2Q_3^2Q_5Q_8-\frac{10}{9}Q_1Q_3^2Q_5Q_7+\frac{5}{3}Q_3Q_5Q_7Q_8-\frac{9}{4}Q_1\times\\
& Q_5Q_8^2+\frac{1}{12}Q_1Q_5\left(Q_7Q_{13}-36Q_3Q_{15}\right),\\
\left[Q_{14},Q_{16}\right]=& 2Q_1^3\left(Q_5^2-2Q_2Q_3^2\right)Q_{13}-12Q_1^3Q_3^3\left(Q_5^2-3Q_2Q_3^2\right)-13Q_1^3Q_3^4Q_6+6Q_1^2Q_2\left(Q_{13}-18Q_3^3\right)Q_8\\
& + 18\left(Q_1^2Q_3Q_5^2-Q_2Q_3^2Q_7\right)Q_8-\frac{1}{12}Q_1\left(Q_6Q_7Q_{13}+7Q_3^2Q_5^2Q_7\right)-\frac{16}{3}Q_1Q_3^2Q_6Q_7+6Q_8Q_9^2\\
& +39Q_1^2Q_3^2Q_6Q_8-\frac{117}{4}Q_1Q_6Q_8^2+17Q_1Q_5^2Q_{15}+2\left(Q_5^2+4Q_3Q_6\right)Q_7Q_8+12Q_1Q_2Q_3^4Q_7\\
& -Q_1Q_2Q_3\left(Q_7Q_{13}-81Q_8^2\right)-\frac{7}{4}Q_1Q_3^2Q_9^2,\\
\left[Q_{14},Q_{17}\right]=&\frac{1}{8}Q_1^2\left(2Q_1Q_3^2Q_9-3Q_8Q_9\right)Q_{13}+\left(\frac{9}{2}Q_1^2Q_3^2Q_8-\frac{3}{2}Q_1^3Q_3^4\right)\left(Q_3Q_9+3Q_{14}\right)+\frac{9}{2}Q_1Q_3Q_{14}Q_{15}\\
& +\frac{1}{2}\left(Q_1Q_3^2-6Q_8\right)Q_9Q_{15}+\frac{3}{2}\left(Q_3Q_7-\frac{27}{4}Q_1Q_8\right)Q_8Q_{14}-\frac{1}{8}Q_1\left(Q_7Q_{13}+8Q_3^3Q_7\right)Q_{14}\\
& +\frac{1}{24}Q_1Q_3Q_7Q_9Q_{13}-\frac{1}{3}Q_1Q_3^4Q_7Q_9+\frac{1}{2}Q_3^2Q_7Q_8Q_9-\frac{27}{8}Q_1Q_3Q_8^2Q_9,\\
\left[Q_{15},Q_{16}\right]=&18Q_1^2Q_3^2\left(Q_1Q_3^2-3Q_8\right)Q_9+\left(2Q_1Q_3^3Q_7-3Q_3Q_7Q_8+\frac{81}{2}Q_1Q_8^2\right)Q_9-12Q_1Q_3^2Q_7Q_{14}\\
& +\frac{3}{2}Q_1Q_7Q_9Q_{13}+18Q_7Q_8Q_{14}+162Q_1Q_{14}Q_{15},\\
\left[Q_{15},Q_{17}\right]=& Q_1^3Q_3^4\left(9Q_{15}-Q_3^2Q_7\right)+\frac{27}{8}\left(Q_8-2Q_1Q_3^2\right)Q_7Q_8^2+\frac{1}{4}Q_1\left(Q_7Q_{13}+81Q_8^2-36Q_3Q-{15}\right)Q_{15}\\
& +\frac{9}{2}Q_1^2Q_3^4Q_7Q_8+\frac{1}{8}\left(Q_8-\frac{2}{3}Q_1Q_3^2\right)Q_7^2Q_{13}+3Q_1Q_3^2\left(Q_3Q_7-9Q_1Q_8\right)Q_{15}-\frac{3}{2}Q_3Q_7Q_8Q_{15},\\
\left[Q_{16},Q_{17}\right]=&27Q_1^5Q_3^6Q_5-\frac{243}{2}Q_1^4Q_3^4Q_5Q_8+\frac{729}{4}Q_1^2\left(Q_1Q_3^2Q_5-\frac{1}{2}Q_5Q_8\right)Q_8^2+\frac{243}{2}Q_1^2Q_3Q_5Q_8Q_{15} \\
&+ 9Q_1Q_3^4\left(Q_5Q_7+4Q_1^2Q_3Q_5\right)Q_7+81Q_1Q_5\left(Q_{15}^2+Q_3Q_7Q_8^2\right) +Q_1^3Q_3^3\left(\frac{9}{2}Q_3Q_{16}-81Q_5Q_{15}\right) \\
&  -108Q_1^2Q_3^3Q_5Q_7Q_8+\frac{9}{8}Q_1^2Q_5\left(Q_7Q_8-\frac{2}{3}Q_1Q_3^2Q_7\right)Q_{13}+\frac{27}{2}Q_5\left(Q_1^3Q_{13}Q_{15}-Q_3^2Q_7^2Q_8\right)\\
& +9Q_5Q_7\left(Q_8-\frac{2}{3}Q_1Q_3^2\right)Q_{15}+\frac{9}{2}Q_3\left(Q_1Q_{15}-Q_7Q_8\right)Q_{16}-\frac{1}{8}Q_1\left(Q_7Q_{13}-81Q_8^2\right)Q_{16}\\
&+3Q_1Q_3^2\left(Q_3Q_7-\frac{9}{2}Q_1Q_8\right)Q_{16}.
\end{split}
\end{equation*}

As only eight among these seventeen polynomials are functionally independent, a certain number of algebraic dependence relations are expected. In contrast to the previous examples, the computation of a complete set of relations for $G_2$ is rather long and cumbersome, ultimately leading to $57$ constraints. 

\begin{equation*}
\begin{split} 
& Q_4Q_9 + 3 Q_5Q_{10} - 2 Q_2Q_{12}=0,\qquad 6Q_2Q_{14}-Q_2Q_3Q_9 + Q_6 Q_9 + Q_5Q_{12}=0,\\
& Q_5^2 Q_7 + 3 Q_9^2 - 12 Q_2 Q_15=0,\qquad 3 Q_1^2 Q_5^3 + Q_5Q_6Q_7 - 3 Q_4Q_{15} + Q_2Q_{16}=0,\\
& 6 Q_ {1}^2  (Q_ {5}^3- Q_ {2} Q_ {3}^2 Q_ {5} ) - 
  3 Q_ {2} Q_ {3} Q_ {5} Q_ {7} + Q_ {5} Q_ {6} Q_ {7} + 
  9 Q_ {1} Q_ {2} Q_ {5} Q_ {8} + 3 Q_ {9} Q_ {12} - 6 Q_ {4} Q_ {15}=0,\\
&  6Q_ {1}^2  Q_ {3}^2 Q_ {5} Q_ {9} + 
  2 Q_ {3} Q_ {5} Q_ {7} Q_ {9} - 9 Q_ {1} Q_ {5} Q_ {8} Q_ {9} + 
  6 Q_ {5} Q_ {7} Q_ {14} + 6 Q_ {12} Q_ {15} + 12 Q_ {2} Q_ {17}=0,\\
 & 6 Q_ {1}^2 Q_ {3}^2 Q_ {5} Q_ {9} + 
  4 Q_ {3} Q_ {5} Q_ {7} Q_ {9} - 9 Q_ {1} Q_ {5} Q_ {8} Q_ {9} - 
  6 Q_ {5} Q_ {7} Q_ {14} - 12 Q_ {12} Q_ {15} + 2 Q_ {9} Q_ {16}=0,\\
&  2 Q_ {2}^2 Q_ {9} - 9 Q_ {2} Q_ {3} Q_ {10} + 3 Q_ {6} Q_ {10} + 
  27 Q_ {1} Q_ {2} Q_ {11} - Q_ {4} Q_ {12}=0,\\
&  2 Q_ {2} Q_ {5} Q_ {9}-Q_ {3} Q_ {4} Q_ {9}   - 
  9 Q_ {3} Q_ {5} Q_ {10} + 27 Q_ {1} Q_ {5} Q_ {11} + 
  2 Q_ {6} Q_ {12} + 6 Q_ {4} Q_ {14}=0,\\
 & 3 Q_ {12}^2-6 Q_ {1}^2 Q_ {3}^2 Q_ {4} Q_ {5} - 
  3 Q_ {1}^2 Q_ {5}^2 Q_ {6} - 3 Q_ {3} Q_ {4} Q_ {5} Q_ {7} - 
  Q_ {2} Q_ {5}^2 Q_ {7} + 9 Q_ {1} Q_ {4} Q_ {5} Q_ {8}  
   - Q_ {4} Q_ {16}=0,\\
&   36 Q_ {1}^2 Q_ {2}^2 Q_ {3}^3 - 12 Q_ {1}^2 Q_ {2} Q_ {3}^2 Q_ {6} + 
  (9 Q_ {2}^2 Q_ {3}^2 - Q_ {2} Q_ {5}^2 - 
  6 Q_ {2} Q_ {3} Q_ {6} + Q_ {6}^2 ) Q_ {7} + 18 Q_ {1} Q_ {2} \left(Q_ {6} - 
  3 Q_ {2} Q_ {3}\right) Q_ {8}  \\
&  - 
  3 Q_ {2} Q_ {9}^2 + 3 Q_ {12}^2 - 3 Q_ {1}^2 Q_ {2}^2 Q_ {13}=0,\\
& 6 Q_ {1}^2 Q_ {3}^2 Q_ {4} Q_ {5} -12 Q_ {1}^2 Q_ {2}^2 Q_ {3}^3 + 
  12 Q_ {1}^2 Q_ {2} Q_ {3} Q_ {5}^2 + 
  (12 Q_ {1}^2 Q_ {2} Q_ {3}^2  - 6 Q_ {1}^2 Q_ {5}^2) Q_ {6} - 
  6 Q_ {2}^2 Q_ {3}^2 Q_ {7} - 
  2 Q_ {6}^2 Q_ {7} + 
  2 Q_ {2} Q_ {5}^2 Q_ {7}\\ 
 & + 2 \left(Q_ {3} Q_ {4} Q_ {5}  + 4 Q_ {2} Q_ {3} Q_ {6}\right) Q_ {7} + 18 Q_ {1} \left(Q_ {2}^2 Q_ {3}  - 
  \frac{1}{2}  Q_ {4} Q_ {5}  -   Q_ {2} Q_ {6}\right) Q_ {8} + 
  54 Q_ {10} Q_ {14}=0,\\
 & 12 Q_ {1}^2 Q_ {2}^2 Q_ {3}^2 - 12 Q_ {1}^2 Q_ {2} Q_ {5}^2 + 
  6 Q_ {2}^2 Q_ {3} Q_ {7} + Q_ {4} Q_ {5} Q_ {7} - 
  2 Q_ {2} Q_ {6} Q_ {7} - 18 Q_ {1} Q_ {2}^2 Q_ {8} - 
  9 Q_ {9} Q_ {10}=0,\\
 & 12 Q_ {1} Q_ {2}^2 Q_ {3}^3 + 2 Q_ {1} Q_ {3}^2 Q_ {4} Q_ {5} + 
 12 Q_ {1} Q_ {2} Q_ {3} Q_ {5}^2 - 4 Q_ {1} Q_ {2} Q_ {3}^2 Q_ {6}-2 Q_ {1} Q_ {5}^2 Q_ {6} - 18 Q_ {2}^2 Q_ {3} Q_ {8} -3 Q_ {4} Q_ {5} Q_ {8}\\
& +6 Q_ {2} Q_ {6} Q_ {8} - 2 Q_ {1} Q_ {2}^2 Q_ {13} + 
  27 Q_ {9} Q_ {11}=0,\\
 & 3( Q_ {3} Q_ {5} Q_ {7}- 
  3 Q_ {1} Q_ {5} Q_ {8}) Q_ {12}  +3 Q_ {1}^2(2 Q_ {3}^2 Q_ {5} Q_ {12} -Q_ {3} Q_ {5}^2 Q_ {9}+ 6 Q_ {5}^2 Q_ {14}) - 
  Q_ {5}^2 Q_ {7} Q_ {9}   + 
  Q_ {12} Q_ {16} + 6 Q_ {4} Q_ {17}=0,\\
&  36 Q_ {1}^2 Q_ {2} Q_ {3}^3 Q_ {9} - 
 12 Q_ {1}^2 Q_ {3}^2 Q_ {6} Q_ {9} + 9 Q_ {2} Q_ {3}^2 Q_ {7} Q_ {9} -Q_ {5}^2 Q_ {7} Q_ {9} - 5 Q_ {3} Q_ {6} Q_ {7} Q_ {9} - 
 54 Q_ {1} Q_ {2} Q_ {3} Q_ {8} Q_ {9} - 6 Q_ {6} Q_ {7} Q_ {14} \\
 &+18 Q_ {1} Q_ {6} Q_ {8} Q_ {9}  -3 Q_ {9}^3  + 6 Q_ {1}^2 Q_ {3}^2 Q_ {5} Q_ {12} + 
 3 Q_ {3} Q_ {5} Q_ {7} Q_ {12} - 9 Q_ {1} Q_ {5} Q_ {8} Q_ {12} -3 Q_ {1}^2 Q_ {2} Q_ {9} Q_ {13}   + 
 2 Q_ {12} Q_ {16}=0,\\
& 6 Q_ {1} Q_ {3}^2 Q_ {10} -2 Q_ {1} Q_ {6} Q_ {9}  - 
  9 Q_ {8} Q_ {10} + 9 Q_ {7} Q_ {11} - 4 Q_ {1} Q_ {5} Q_ {12}=0,\\
 & 6 Q_ {1}^2 Q_ {2} Q_ {3}^2 Q_ {9} - 6 Q_ {1}^2 Q_ {5}^2 Q_ {9} + 
  3 Q_ {2} Q_ {3} Q_ {7} Q_ {9} - Q_ {6} Q_ {7} Q_ {9} - 
  9 Q_ {1} Q_ {2} Q_ {8} Q_ {9} + Q_ {5} Q_ {7} Q_ {12} - 
  18 Q_ {10} Q_ {15}=0,\\
 & 4 Q_ {5}^2 Q_ {9} + 4 Q_ {3} Q_ {6} Q_ {9} + 
  54 Q_ {1} Q_ {3}^2 Q_ {11} - 81 Q_ {8} Q_ {11} + 
  12 Q_ {3} Q_ {5} Q_ {12} - 3 Q_ {10} Q_ {13} + 12 Q_ {6} Q_ {14}=0,\\
 & 6 Q_ {1} Q_ {2} Q_ {3}^3 Q_ {9} + 4 Q_ {1} Q_ {3} Q_ {5}^2 Q_ {9} - 
 2 Q_ {1} Q_ {3}^2 Q_ {6} Q_ {9} - 9 Q_ {2} Q_ {3} Q_ {8} Q_ {9} + 
 3 Q_ {6} Q_ {8} Q_ {9} + 2 Q_ {1} Q_ {3}^2 Q_ {5} Q_ {12}-3 Q_ {5} Q_ {8} Q_ {12}\\
&  -Q_ {1} Q_ {2} Q_ {9} Q_ {13} + 12 Q_ {1} Q_ {5}^2 Q_ {14} + 
  54 Q_ {11} Q_ {15}=0,\\
  & 12 Q_ {1}^2 Q_ {2} Q_ {3}^4 - 18 Q_ {1}^2 Q_ {3}^2 Q_ {5}^2 - 
  9 Q_ {3} Q_ {5}^2 Q_ {7} - 36 Q_ {1} Q_ {2} Q_ {3}^2 Q_ {8} + 
  27 Q_ {1} Q_ {5}^2 Q_ {8} + 27 Q_ {2} Q_ {8}^2 - 3 Q_ {3} Q_ {9}^2 +
   Q_ {2} Q_ {7} Q_ {13}\\
   & + 18 Q_ {9} Q_ {14} - 2 Q_ {5} Q_ {16}=0,\\
   & 12 Q_ {1}^2 Q_ {2} Q_ {3}^4 - 12 Q_ {1}^2 Q_ {3}^2 Q_ {5}^2 - 
  6 Q_ {3} Q_ {5}^2 Q_ {7} - 36 Q_ {1} Q_ {2} Q_ {3}^2 Q_ {8} + 
  18 Q_ {1} Q_ {5}^2 Q_ {8} + 27 Q_ {2} Q_ {8}^2 - 6 Q_ {3} Q_ {9}^2 +
   Q_ {2} Q_ {7} Q_ {13} \\
   & + 36 Q_ {9} Q_ {14} + 12 Q_ {6} Q_ {15}=0,\\
   & 12 Q_ {1}^2 Q_ {3}^4 Q_ {9} - 36 Q_ {1} Q_ {3}^2 Q_ {8} Q_ {9} + 
  27 Q_ {8}^2 Q_ {9} + Q_ {7} Q_ {9} Q_ {13} - 
  12 Q_ {3} Q_ {9} Q_ {15} + 72 Q_ {14} Q_ {15} + 24 Q_ {5} Q_ {17}=0,\\
  & 36 Q_ {1} Q_ {3}^2 Q_ {4} Q_ {8} - 12 Q_ {1}^2 Q_ {3}^4 Q_ {4} - 
  144 Q_ {1}^2 Q_ {2} Q_ {3}^3 Q_ {5} + 
  48 Q_ {1}^2 Q_ {3}^2 Q_ {5} Q_ {6} - 
  36 Q_ {2} Q_ {3}^2 Q_ {5} Q_ {7} + 24 Q_ {3} Q_ {5} Q_ {6} Q_ {7} + 
  12 Q_ {5} Q_ {9}^2\\
  & + 
  216 Q_ {1} Q_ {2} Q_ {3} Q_ {5} Q_ {8}- 
  72 Q_ {1} Q_ {5} Q_ {6} Q_ {8} - 27 Q_ {4} Q_ {8}^2   + 12 Q_ {1}^2 Q_ {2} Q_ {5} Q_ {13} - 
  Q_ {4} Q_ {7} Q_ {13} + 4 Q_ {6} Q_ {16}=0,\\
  & 12 Q_ {1}^2 Q_ {3}^4 Q_ {4} + 72 Q_ {1}^2 Q_ {2} Q_ {3}^3 Q_ {5} - 
 12 Q_ {1}^2 Q_ {3}^2 Q_ {5} Q_ {6} + 2 \left(Q_ {5}^3 +  
 9 Q_ {2} Q_ {3}^2 Q_ {5}-3Q_ {3} Q_ {5} Q_ {6}\right) Q_ {7} - 36 Q_ {1} Q_ {3}^2 Q_ {4} Q_ {8}-6 Q_ {5} Q_ {9}^2\\
 &  -108 Q_ {1} Q_ {2} Q_ {3} Q_ {5} Q_ {8} + 
  18 Q_ {1} Q_ {5} Q_ {6} Q_ {8} + 27 Q_ {4} Q_ {8}^2 - 
  6 Q_ {3} Q_ {9} Q_ {12} - 6 Q_ {1}^2 Q_ {2} Q_ {5} Q_ {13} + 
  Q_ {4} Q_ {7} Q_ {13} + 36 Q_ {12} Q_ {14}=0,\\
  & 12 Q_ {3} Q_ {5} Q_ {7} Q_ {14} - 
 (36 Q_ {1}^2 Q_ {3}^3 Q_ {5}  - 
 8 Q_ {3}^2 Q_ {5} Q_ {7}  + 
 54 Q_ {1} Q_ {3} Q_ {5} Q_ {8} )Q_ {9} + 
 3 Q_ {1}^2 Q_ {5} Q_ {9} Q_ {13} + 
 36 Q_ {1}^2 Q_ {3}^2 Q_ {5} Q_ {14}+12 Q_ {14} Q_ {16}\\
 &  -54 Q_ {1} Q_ {5} Q_ {8} Q_ {14} + 12 Q_ {5} Q_ {9} Q_ {15} - 
  12 Q_ {3} Q_ {12} Q_ {15} - 24 Q_ {6} Q_ {17}=0,\\
  & 24 Q_ {1}^2 Q_ {3}^3 Q_ {5} Q_ {9} + 
 4 Q_ {3}^2 Q_ {5} Q_ {7} Q_ {9} - 
 36 Q_ {1} Q_ {3} Q_ {5} Q_ {8} Q_ {9} + 
 12 Q_ {1}^2 Q_ {3}^4 Q_ {12} - 36 Q_ {1} Q_ {3}^2 Q_ {8} Q_ {12} +(Q_ {7} Q_ {12} -3 Q_ {1}^2 Q_ {5} Q_ {9} )Q_ {13}\\
 & + 
 27 Q_ {8}^2 Q_ {12} + \left(
  108 Q_ {1}^2 Q_ {3}^2 Q_ {5}  + 
  48 Q_ {3} Q_ {5} Q_ {7}  - 
  162 Q_ {1} Q_ {5} Q_ {8}\right) Q_ {14} - 12 Q_ {5} Q_ {9} Q_ {15} - 
  12 Q_ {3} Q_ {12} Q_ {15} + 12 Q_ {14} Q_ {16}=0,\\
  & 108 Q_ {3} Q_ {5} Q_ {7} Q_ {15} - 
 12 Q_ {1}^2 Q_ {3}^4 Q_ {5} Q_ {7} + 
 36 Q_ {1} Q_ {3}^2 Q_ {5} Q_ {7} Q_ {8} - 27 Q_ {5} Q_ {7} Q_ {8}^2 -
  Q_ {5} Q_ {7}^2 Q_ {13}+216 Q_ {1}^2 Q_ {3}^2 Q_ {5} Q_ {15}  +24 Q_ {15} Q_ {16} \\
 & - 
 324 Q_ {1} Q_ {5} Q_ {8} Q_ {15}+ 72 Q_ {9} Q_ {17}=0,\\
 & Q_ {4}^2 Q_ {7}-12 Q_ {1}^2 Q_ {2} Q_ {4} Q_ {5} - 12 Q_ {1}^2 Q_ {2}^2 Q_ {6} - 
  4 Q_ {2}^3 Q_ {7}   + 27 Q_ {10}^2=0,\\
  & 8 Q_ {1} Q_ {2}^3 Q_ {3}^2 - 2 Q_ {1} Q_ {3}^2 Q_ {4}^2 - 
 12 Q_ {1} Q_ {2} Q_ {3} Q_ {4} Q_ {5} - 8 Q_ {1} Q_ {2}^2 Q_ {5}^2 - 
 12 Q_ {1} Q_ {2}^2 Q_ {3} Q_ {6} + 2 Q_ {1} Q_ {4} Q_ {5} Q_ {6}+4 Q_ {1} Q_ {2} Q_ {6}^2 - 12 Q_ {2}^3 Q_ {8}\\
 & +3 Q_ {4}^2 Q_ {8} + 81 Q_ {10} Q_ {11}=0,\\
& Q_ {4} Q_ {6} Q_ {7} - 6 Q_ {1}^2 Q_ {2} Q_ {3}^2 Q_ {4} - 
  6 Q_ {1}^2 Q_ {2} Q_ {5} Q_ {6} - 3 Q_ {2} Q_ {3} Q_ {4} Q_ {7} - 
  2 Q_ {2}^2 Q_ {5} Q_ {7} + 9 Q_ {1} Q_ {2} Q_ {4} Q_ {8} + 
  9 Q_ {10} Q_ {12}=0,\\
  & 9 Q_ {3} Q_ {5} Q_ {7} Q_ {10} - 6 Q_ {1}^2 Q_ {3}^2 Q_ {4} Q_ {9} - 
  6 Q_ {1}^2 Q_ {5} Q_ {6} Q_ {9} - 3 Q_ {3} Q_ {4} Q_ {7} Q_ {9} - 
  2 Q_ {2} Q_ {5} Q_ {7} Q_ {9} + 9 Q_ {1} Q_ {4} Q_ {8} Q_ {9} + 
  18 Q_ {1}^2 Q_ {3}^2 Q_ {5} Q_ {10}\\
  &  - 
  27 Q_ {1} Q_ {5} Q_ {8} Q_ {10} + 2 Q_ {6} Q_ {7} Q_ {12} + 
  6 Q_ {10} Q_ {16}=0,\\
\end{split}
\end{equation*}

\begin{equation*}
\begin{split} 
  &12 Q_ {1}^2 (Q_ {3}^2 Q_ {5} Q_ {6} 
 - Q_ {3}^4 Q_ {4}  - 
6 Q_ {2} Q_ {3}^3 Q_ {5}) Q_ {7} + -18 Q_ {2} Q_ {3}^2 Q_ {5} Q_ {7}^2 - 2 Q_ {5}^3 Q_ {7}^2 + 
 6 Q_ {3} Q_ {5} Q_ {6} Q_ {7}^2 + 
 36 Q_ {1} Q_ {3}^2 Q_ {4} Q_ {7} Q_ {8}\\
 &  +108 Q_ {1} Q_ {2} Q_ {3} Q_ {5} Q_ {7} Q_ {8} - 
 18 Q_ {1} Q_ {5} Q_ {6} Q_ {7} Q_ {8} - 27 Q_ {4} Q_ {7} Q_ {8}^2 + 
 36 Q_ {1}^2 Q_ {3} Q_ {5} Q_ {9}^2 + 18 Q_ {5} Q_ {7} Q_ {9}^2 + 
 36 Q_ {1}^2 Q_ {3}^2 Q_ {9} Q_ {12}\\
 & +18 Q_ {3} Q_ {7} Q_ {9} Q_ {12} - 54 Q_ {1} Q_ {8} Q_ {9} Q_ {12} + 
  6 Q_ {1}^2 Q_ {2} Q_ {5} Q_ {7} Q_ {13} - Q_ {4} Q_ {7}^2 Q_ {13} - 
  216 Q_ {1}^2 Q_ {5} Q_ {9} Q_ {14} + 216 Q_ {10} Q_ {17}=0,\\
  & 12 Q_ {3} Q_ {4} Q_ {5} Q_ {6} - 36 Q_ {2} Q_ {3}^2 Q_ {4} Q_ {5} - 
  48 Q_ {2}^2 Q_ {3} Q_ {5}^2 - 4 Q_ {4} Q_ {5}^3 - 
  (36 Q_ {2}^2 Q_ {3}^2  + 12 Q_ {2} Q_ {5}^2) Q_ {6} + 
  24 Q_ {2} Q_ {3} Q_ {6}^2 - 4 Q_ {6}^3 + 4 Q_ {2}^3 Q_ {13} \\
  & -  Q_ {4}^2 Q_ {13} + 729 Q_ {11}^2=0,\\
  & 6 Q_ {1} Q_ {2} Q_ {3}^3 Q_ {4} + 4 Q_ {1} Q_ {2}^2 Q_ {3}^2 Q_ {5} - 
  4 Q_ {1} Q_ {2} Q_ {5}^3 - 2 Q_ {1} \left(Q_ {3}^2 Q_ {4}  +3 Q_ {2} Q_ {3} Q_ {5}\right) Q_ {6} + 2 Q_ {1} Q_ {5} Q_ {6}^2 - 
  9 Q_ {2} Q_ {3} Q_ {4} Q_ {8} - 6 Q_ {2}^2 Q_ {5} Q_ {8}\\
  & + 
  3 Q_ {4} Q_ {6} Q_ {8} - Q_ {1} Q_ {2} Q_ {4} Q_ {13} + 
  27 Q_ {11} Q_ {12}=0,\\
  & 12 Q_ {1} Q_ {2}^2 Q_ {3}^4 - 4 Q_ {1} Q_ {3}^3 Q_ {4} Q_ {5} + 
  8 Q_ {1} Q_ {2} Q_ {3}^2 Q_ {5}^2 + 4 Q_ {1} Q_ {5}^4 - 
  16 Q_ {1} Q_ {2} Q_ {3}^3 Q_ {6} - 
  8 Q_ {1} Q_ {3} Q_ {5}^2 Q_ {6} + 4 Q_ {1} Q_ {3}^2 Q_ {6}^2 - 
  18 Q_ {2}^2 Q_ {3}^2 Q_ {8} \\
  & + 6 \left(Q_ {3} Q_ {4} Q_ {5} + 
   Q_ {2} Q_ {5}^2  + 4 Q_ {2} Q_ {3} Q_ {6}  - 
 6 Q_ {6}^2\right)Q_ {8} - 2 Q_ {1} Q_ {2}^2 Q_ {3} Q_ {13} + 
  Q_ {1} (Q_ {4} Q_ {5}  + 2  Q_ {2} Q_ {6})Q_ {13} + 
  162 Q_ {11} Q_ {14}=0,\\
  & 6 Q_ {1} Q_ {3}^3 Q_ {4} Q_ {9} + 
 4 Q_ {1} Q_ {2} Q_ {3}^2 Q_ {5} Q_ {9} + 4 Q_ {1} Q_ {5}^3 Q_ {9} + 
 12 Q_ {1} Q_ {3} Q_ {5} Q_ {6} Q_ {9} - 
 9 Q_ {3} Q_ {4} Q_ {8} Q_ {9} - 6 Q_ {2} Q_ {5} Q_ {8} Q_ {9}+6 Q_ {6} Q_ {8} Q_ {12}\\
 & -18 Q_ {1} Q_ {3}^3 Q_ {5} Q_ {10} + 
 27 Q_ {3} Q_ {5} Q_ {8} Q_ {10} + 
 162 Q_ {1}^2 Q_ {3}^2 Q_ {5} Q_ {11} - 
 243 Q_ {1} Q_ {5} Q_ {8} Q_ {11} + 
 36 Q_ {1} Q_ {3} Q_ {5}^2 Q_ {12} - 
 4 Q_ {1} Q_ {3}^2 Q_ {6} Q_ {12} \\
 & - Q_ {1} Q_ {4} Q_ {9} Q_ {13}-6 Q_ {1} Q_ {5} Q_ {10} Q_ {13} + 18 Q_ {11} Q_ {16}=0,\\
 & Q_ {6} Q_ {7} Q_ {13} - 36 Q_ {1}^2 Q_ {3}^3 Q_ {5}^2 + 
  12 Q_ {1}^2 Q_ {3}^4 Q_ {6} - 9 Q_ {3}^2 Q_ {5}^2 Q_ {7} + 
   Q_ {1} \left(54Q_ {3} Q_ {5}^2  - 
  36  Q_ {3}^2 Q_ {6} \right)Q_ {8}  + 27 Q_ {6} Q_ {8}^2- 
  3 Q_ {3}^2 Q_ {9}^2 + 3 Q_ {1}^2 Q_ {5}^2 Q_ {13} \\
  &+ 
  36 Q_ {3} Q_ {9} Q_ {14} + 12 Q_ {5}^2 Q_ {15} - 108 Q_ {14}^2=0,\\
  & 24 Q_ {1}^3 (Q_ {3}^6 Q_ {4} + 6 Q_ {2} Q_ {3}^5 Q_ {5} + 
6 Q_ {3}^3 Q_ {5}^3 - 3  Q_ {3}^4 Q_ {5} Q_ {6})+Q_ {1} (36Q_ {2} Q_ {3}^4 Q_ {5} + 
 40 Q_ {3}^2 Q_ {5}^3-12 Q_ {3}^3 Q_ {5} Q_ {6}) Q_ {7}-48 Q_ {1} Q_ {5}^3 Q_ {15}\\
 & -108 Q_ {1}^2 (Q_ {3}^4 Q_ {4} +4 Q_ {2} Q_ {3}^3 Q_ {5} 2 Q_ {3} Q_ {5}^3 -2 Q_ {3}^2 Q_ {5} Q_ {6}) Q_ {8} - 6 (Q_ {5}^3 -3 Q_ {3} Q_ {5} Q_ {6}+9Q_ {2} Q_ {3}^2 Q_ {5}) Q_ {7} Q_ {8} +162 Q_ {1} Q_ {3}^2 Q_ {4} Q_ {8}^2\\
 & +\left(324 Q_ {1} Q_ {2} Q_ {3} Q_ {5}  - 
 162 Q_ {1} Q_ {5} Q_ {6}\right) Q_ {8}^2 - 81 Q_ {4} Q_ {8}^3 + 
 54 Q_ {5} Q_ {8} Q_ {9}^2 - \left(36 Q_ {1} Q_ {3}^3 Q_ {9} +54 Q_ {3} Q_ {8} Q_ {9}\right) Q_ {12} - 
 12 Q_ {1}^3 Q_ {2} Q_ {3}^2 Q_ {5} Q_ {13}\\
 &  -12 Q_ {1}^3 Q_ {5}^3 Q_ {13} + 
 2 Q_ {1} Q_ {3}^2 Q_ {4} Q_ {7} Q_ {13} - 
 4 Q_ {1} Q_ {5} Q_ {6} Q_ {7} Q_ {13} + 
 18 Q_ {1}^2 Q_ {2} Q_ {5} Q_ {8} Q_ {13} -3 Q_ {4} Q_ {7} Q_ {8} Q_ {13} + 6 Q_ {1} Q_ {9} Q_ {12} Q_ {13} \\
 & - 
  216 Q_ {1} Q_ {3} Q_ {5} Q_ {9} Q_ {14} + 648 Q_ {11} Q_ {17}=0,\\
  &  
 144 Q_ {1}^2 Q_ {2} (Q_ {3}^5 Q_ {7}+2Q_ {1}^2  Q_ {3}^6)- 
 324 Q_ {1}^2 Q_ {3}^3 Q_ {5}^2( Q_ {7} + Q_ {1}^2 Q_ {3})+324 \left(Q_ {3} Q_ {7} Q_ {9}   -3Q_ {1} Q_ {8} Q_ {9}\right) Q_ {14} - 36 Q_ {9}^2 Q_ {15} -9 Q_ {1}^2 Q_ {9}^2 Q_ {13} \\
 & +12 Q_ {1}^2 Q_ {3}^4 Q_ {6} Q_ {7} - 81 Q_ {3}^2 Q_ {5}^2 Q_ {7}^2 + 
 324 Q_ {1}^3 (3Q_ {3}^2 Q_ {5}^2 - 
4 Q_ {2} Q_ {3}^4) Q_ {8}  +( 
 486 Q_ {1} Q_ {3} Q_ {5}^2- 
 432 Q_ {1} Q_ {2} Q_ {3}^3 ) Q_ {7} Q_ {8}-729 Q_ {1}^2 Q_ {5}^2 Q_ {8}^2\\
 & -36 Q_ {1} Q_ {3}^2 Q_ {6} Q_ {7} Q_ {8} + 
 1944 Q_ {1}^2 Q_ {2} Q_ {3}^2 Q_ {8}^2 + 27 (Q_ {6} + 
 12 Q_ {2} Q_ {3}) Q_ {7} Q_ {8}^2 -972 Q_ {1} Q_ {2} Q_ {8}^3 - 27 Q_ {3}^2 Q_ {7} Q_ {9}^2 + 
 24 Q_ {1}^2 Q_ {2} Q_ {3}^2 Q_ {7} Q_ {13}\\
 &  +12 Q_ {2} Q_ {3} Q_ {7}^2 Q_ {13} + Q_ {6} Q_ {7}^2 Q_ {13} - 
 36 Q_ {1} Q_ {2} Q_ {7} Q_ {8} Q_ {13} + 
 648 Q_ ({1}^2) Q_ ({3}^2) Q_ {9} Q_ {14} + 4 Q_ {16}^2=0,\\
 & 12 Q_ {1}^2 Q_ {3}^3 (3Q_ {5}^2  - 
 12 Q_ {3} Q_ {6}) Q_ {7} +9 Q_ {3}^2 Q_ {5}^2 Q_ {7}^2 - 
 54 Q_ {1} Q_ {3} Q_ {5}^2 Q_ {7} Q_ {8} + 
 36 Q_ {1} Q_ {3}^2 Q_ {6} Q_ {7} Q_ {8} - 27 Q_ {6} Q_ {7} Q_ {8}^2 - 9 (Q_ {3}^2 Q_ {7} +8Q_ {1}^2 Q_ {3}^3) Q_ {9}^2 \\
 & +108 Q_ {1} Q_ {3} Q_ {8} Q_ {9}^2 - 
 3 Q_ {1}^2 Q_ {5}^2 Q_ {7} Q_ {13} - Q_ {6} Q_ {7}^2 Q_ {13} + 
 9 Q_ {1}^2 Q_ {9}^2 Q_ {13} - 216 Q_ {1}^2 Q_ {3}^2 Q_ {9} Q_ {14} - 
 108 Q_ {3} Q_ {7} Q_ {9} Q_ {14} + 324 Q_ {1} Q_ {8} Q_ {9} Q_ {14}\\
 & -12 Q_ {5}^2 Q_ {7} Q_ {15} + 36 Q_ {9}^2 Q_ {15} + 72 Q_ {12} Q_ {17}=0,\\
 & 144 Q_ {1} Q_ {3}^3 Q_ {5} Q_ {7} Q_ {8} - 
 486 Q_ {1}^2 Q_ {3}^2 Q_ {5} Q_ {8}^2 - 
 108 Q_ {3} Q_ {5} Q_ {7} Q_ {8}^2 + 243 Q_ {1} Q_ {5} Q_ {8}^3-72 Q_ {1}^4 Q_ {3}^6 Q_ {5} - 48 Q_ {1}^2 Q_ {3}^5 Q_ {5} Q_ {7} + 
 324 Q_ {1}^3 Q_ {3}^4 Q_ {5} Q_ {8}\\
 & -6 Q_ {1}^2 Q_ {3}^2 Q_ {5} Q_ {7} Q_ {13} - 
 4 Q_ {3} Q_ {5} Q_ {7}^2 Q_ {13} + 
 9 Q_ {1} Q_ {5} Q_ {7} Q_ {8} Q_ {13} - 
 216 Q_ {1}^2 Q_ {3}^3 Q_ {5} Q_ {15}+324 Q_ {1} Q_ {3} Q_ {5} Q_ {8} Q_ {15} + 
 36 Q_ {1}^2 Q_ {5} Q_ {13} Q_ {15} \\
 & + 144 Q_ {5} Q_ {15}^2-24 Q_ {1}^2 Q_ {3}^4 Q_ {16} + 72 Q_ {1} Q_ {3}^2 Q_ {8} Q_ {16} - 
  54 Q_ {8}^2 Q_ {16} - 2 Q_ {7} Q_ {13} Q_ {16} + 
  24 Q_ {3} Q_ {15} Q_ {16} + 432 Q_ {14} Q_ {17}=0,\\
  & 9 Q_ {1} Q_ {7} Q_ {8} Q_ {9} Q_ {13}+243 Q_ {1} Q_ {8}^3 Q_ {9} - 
 6 Q_ {1}^2 Q_ {3}^2 Q_ {7} Q_ {9} Q_ {13} - 
 4 Q_ {3} Q_ {7}^2 Q_ {9} Q_ {13} + 
 72 Q_ {1}^2 Q_ {3}^4 Q_ {7} Q_ {14} - 
 216 Q_ {1} Q_ {3}^2 Q_ {7} Q_ {8} Q_ {14}\\
 & -72 Q_ {1}^4 Q_ {3}^6 Q_ {9} - 48 Q_ {1}^2 Q_ {3}^5 Q_ {7} Q_ {9} + 
 (324 Q_ {1}^3 Q_ {3}^4  + 
 144 Q_ {1} Q_ {3}^3 Q_ {7}) Q_ {8} Q_ {9} - 
 486 Q_ {1}^2 Q_ {3}^2 Q_ {8}^2 Q_ {9} - 
 108 Q_ {3} Q_ {7} Q_ {8}^2 Q_ {9} + 162 Q_ {7} Q_ {8}^2 Q_ {14}\\
 & +6 Q_ {7}^2 Q_ {13} Q_ {14} - 216 Q_ {1}^2 Q_ {3}^3 Q_ {9} Q_ {15} + 
 324 Q_ {1} Q_ {3} Q_ {8} Q_ {9} Q_ {15} + 
 36 Q_ {1}^2 Q_ {9} Q_ {13} Q_ {15} - 
 1296 Q_ {1}^2 Q_ {3}^2 Q_ {14} Q_ {15} - 
 648 Q_ {3} Q_ {7} Q_ {14} Q_ {15}\\
 & +1944 Q_ {1} Q_ {8} Q_ {14} Q_ {15} + 144 Q_ {9} Q_ {15}^2 + 
  48 Q_ {16} Q_ {17}=0,\\
  & 144 Q_ {1}^4 Q_ {3}^8 Q_ {7} - 864 Q_ {1}^3 Q_ {3}^6 Q_ {7} Q_ {8} + 
 1944 Q_ {1}^2 Q_ {3}^4 Q_ {7} Q_ {8}^2 - 
 1944 Q_ {1} Q_ {3}^2 Q_ {7} Q_ {8}^3 + 729 Q_ {7} Q_ {8}^4 + 
 24 Q_ {1}^2 Q_ {3}^4 Q_ {7}^2 Q_ {13} - 
 72 Q_ {1} Q_ {3}^2 Q_ {7}^2 Q_ {8} Q_ {13}\\
 & +54 Q_ {7}^2 Q_ {8}^2 Q_ {13} + Q_ {7}^3 Q_ {13}^2 - 
 1728 Q_ {1}^4 Q_ {3}^6 Q_ {15} - 
 864 Q_ {1}^2 Q_ {3}^5 Q_ {7} Q_ {15} + 
 7776 Q_ {1}^3 Q_ {3}^4 Q_ {8} Q_ {15} + 
 2592 Q_ {1} Q_ {3}^3 Q_ {7} Q_ {8} Q_ {15}-1728 Q_ {15}^3\\
 & -11664 Q_ {1}^2 Q_ {3}^2 Q_ {8}^2 Q_ {15} - 
 1944 Q_ {3} Q_ {7} Q_ {8}^2 Q_ {15} + 5832 Q_ {1} Q_ {8}^3 Q_ {15} - 
 72 (Q_ {3} Q_ {7}^2 + 
2 Q_ {1}^2 Q_ {3}^2 Q_ {7} -3 
 216 Q_ {1} Q_ {7} Q_ {8}) Q_ {13} Q_ {15}\\
 & +5184 Q_ {1}^2 Q_ {3}^3 Q_ {15}^2 + 1296 Q_ {3}^2 Q_ {7} Q_ {15}^2 - 
  7776 Q_ {1} Q_ {3} Q_ {8} Q_ {15}^2 - 
  432 Q_ {1}^2 Q_ {13} Q_ {15}^2 + 1728 Q_ {17}^2=0,\\
  & -4 Q_ {2}^3 Q_ {9} + Q_ {4}^2 Q_ {9} + 18 Q_ {2}^2 Q_ {3} Q_ {10} + 
  3 Q_ {4} Q_ {5} Q_ {10} - 6 Q_ {2} Q_ {6} Q_ {10} - 
  54 Q_ {1} Q_ {2}^2 Q_ {11}=0,\\
  & -6 Q_ {3} Q_ {4} Q_ {5} Q_ {9} - 4 Q_ {2} Q_ {5}^2 Q_ {9} - 
  6 Q_ {2} Q_ {3} Q_ {6} Q_ {9} + 2 Q_ {6}^2 Q_ {9} - 
  18 Q_ {3} Q_ {5}^2 Q_ {10} - 54 Q_ {1} Q_ {2} Q_ {3}^2 Q_ {11} + 
  81 Q_ {2} Q_ {8} Q_ {11} \\
  & + 2 Q_ {5} Q_ {6} Q_ {12} + 
  3 Q_ {2} Q_ {10} Q_ {13}=0,\\
  & 4 Q_ {2} Q_ {5} Q_ {6} Q_ {9} + 108 Q_ {2} Q_ {3}^2 Q_ {5} Q_ {10} - 
 54 Q_ {3} Q_ {5} Q_ {6} Q_ {10} - 54 Q_ {1} Q_ {3}^2 Q_ {4} Q_ {11}-24 Q_ {2}^2 Q_ {3} Q_ {5} Q_ {9} - 4 Q_ {4} Q_ {5}^2 Q_ {9} - 
 6 Q_ {3} Q_ {4} Q_ {6} Q_ {9} \\
 & -  324 Q_ {1} Q_ {2} Q_ {3} Q_ {5} Q_ {11}+54 Q_ {1} Q_ {5} Q_ {6} Q_ {11} + 81 Q_ {4} Q_ {8} Q_ {11} + 
  4 Q_ {6}^2 Q_ {12} + 3 Q_ {4} Q_ {10} Q_ {13}=0,\\
  & 36 Q_ {1} Q_ {2}^2 Q_ {3}^2 Q_ {8} - 
 18 Q_ {1} Q_ {2} Q_ {5}^2 Q_ {8} - 27 Q_ {2}^2 Q_ {8}^2 -12 Q_ {1}^2 Q_ {2}^2 Q_ {3}^4 + 
 12 Q_ {1}^2 Q_ {2} Q_ {3}^2 Q_ {5}^2 + 
 6 Q_ {2} Q_ {3} Q_ {5}^2 Q_ {7} -Q_ {5}^2 Q_ {6} Q_ {7}+ 
 3 Q_ {6} Q_ {9}^2\\
 & -Q_ {2}^2 Q_ {7} Q_ {13} + 6 Q_ {5} Q_ {9} Q_ {12}=0,\\
 \end{split}
\end{equation*}

\begin{equation*}
\begin{split} 
&  18 Q_ {2} Q_ {3} Q_ {5} Q_ {6} Q_ {7} - 2 Q_ {5} Q_ {6}^2 Q_ {7} + 
 36 Q_ {1} Q_ {2} Q_ {3}^2 Q_ {4} Q_ {8}-12 Q_ {1}^2 Q_ {2} Q_ {3}^4 Q_ {4} - 
 144 Q_ {1}^2 Q_ {2}^2 Q_ {3}^3 Q_ {5} + 
 36 Q_ {1}^2 Q_ {2} Q_ {3}^2 Q_ {5} Q_ {6}\\
 & -  36 Q_ ({2}^2) Q_ ({3}^2) Q_ {5} Q_ {7} +216 Q_ {1} Q_ {2}^2 Q_ {3} Q_ {5} Q_ {8} - 
  54 Q_ {1} Q_ {2} Q_ {5} Q_ {6} Q_ {8} - 27 Q_ {2} Q_ {4} Q_ {8}^2 + 
  12 Q_ {2} Q_ {5} Q_ {9}^2 + 12 Q_ {1}^2 Q_ {2}^2 Q_ {5} Q_ {13}\\
  & - 
  Q_ {2} Q_ {4} Q_ {7} Q_ {13} + 6 Q_ {6} Q_ {9} Q_ {12}=0,\\
  & 12 Q_ {1}^2 (Q_ {3}^4 Q_ {4}^2 -4  Q_ {2}^3 Q_ {3}^4 + 
 12 Q_ {2} Q_ {3}^3 Q_ {4} Q_ {5} + 
 8 Q_ {2}^2 Q_ {3}^2 Q_ {5}^2 - 4 Q_ {2} Q_ {5}^4 +12 Q_ {2}^2 Q_ {3}^3 Q_ {6}
 -2 Q_ {3}^2 Q_ {4} Q_ {5} Q_ {6} - 
 4  Q_ {2} Q_ {3}^2 Q_ {6}^2 + Q_ {5}^2 Q_ {6}^2)\\
 & +36 Q_ {2} Q_ {3}^2 Q_ {4} Q_ {5} Q_ {7} + 
 48 Q_ {2}^2 Q_ {3} Q_ {5}^2 Q_ {7} + 4 Q_ {4} Q_ {5}^3 Q_ {7} + 
 36 Q_ {2}^2 Q_ {3}^2 Q_ {6} Q_ {7} - 
 12 Q_ {2} Q_ {5}^2 Q_ {6} Q_ {7} - 24 Q_ {2} Q_ {3} Q_ {6}^2 Q_ {7}+4 Q_ {6}^3 Q_ {7}\\
 & +144 Q_ {1} Q_ {2}^3 Q_ {3}^2 Q_ {8} - 
 36 Q_ {1} Q_ {3}^2 Q_ {4}^2 Q_ {8} - 
 216 Q_ {1} Q_ {2} Q_ {3} Q_ {4} Q_ {5} Q_ {8} - 
 144 Q_ {1} Q_ {2}^2 Q_ {5}^2 Q_ {8} - 
 216 Q_ {1} Q_ {2}^2 Q_ {3} Q_ {6} Q_ {8}+27 Q_ {4}^2 Q_ {8}^2\\
 & +36 Q_ {1} Q_ {4} Q_ {5} Q_ {6} Q_ {8} + 
  72 Q_ {1} Q_ {2} Q_ {6}^2 Q_ {8} - 108 Q_ {2}^3 Q_ {8}^2 - 
  12 Q_ {1}^2 Q_ {2} Q_ {4} Q_ {5} Q_ {13} - 
  12 Q_ {1}^2 Q_ {2}^2 Q_ {6} Q_ {13} - 4 Q_ {2}^3 Q_ {7} Q_ {13} + 
  Q_ {4}^2 Q_ {7} Q_ {13}\\
  & - 12 Q_ {3} Q_ {4} Q_ {5} Q_ {6} Q_ {7}=0,\\
  & 12 Q_ {1}^2 (Q_ {2}^2 Q_ {5}^2 Q_ {6}+12 Q_ {2}^4 Q_ {3}^3 + 
 2  Q_ {2}^2 Q_ {3}^2 Q_ {4} Q_ {5} - 
   Q_ {2} Q_ {4} Q_ {5}^3 - 4 Q_ {2}^3 Q_ {3}^2 Q_ {6})+36 Q_ {2}^4 Q_ {3}^2 Q_ {7} + 
 12 Q_ {2}^2 Q_ {3} Q_ {4} Q_ {5} Q_ {7} + Q_ {4}^2 Q_ {5}^2 Q_ {7}\\
 & -24 Q_ {2}^3 Q_ {3} Q_ {6} Q_ {7} + 4 Q_ {2}^2 Q_ {6}^2 Q_ {7} - 
 216 Q_ {1} Q_ {2}^4 Q_ {3} Q_ {8} - 
 36 Q_ {1} Q_ {2}^2 Q_ {4} Q_ {5} Q_ {8} + 
 72 Q_ {1} Q_ {2}^3 Q_ {6} Q_ {8}-12 Q_ {2}^3 Q_ {9}^2 -12 Q_ {1}^2 Q_ {2}^4 Q_ {13} \\
 & + 3 Q_ {4}^2 Q_ {9}^2- 4 Q_ {2} Q_ {4} Q_ {5} Q_ {6} Q_ {7}=0,\\
 & 36 Q_ {2} Q_ {3}^2 Q_ {4} Q_ {5} Q_ {9} + 
 48 Q_ {2}^2 Q_ {3} Q_ {5}^2 Q_ {9} + 4 Q_ {4} Q_ {5}^3 Q_ {9} + 
 36 Q_ {2}^2 Q_ {3}^2 Q_ {6} Q_ {9} - 
 12 Q_ {2} Q_ {5}^2 Q_ {6} Q_ {9} - 
 24 Q_ {2} Q_ {3} Q_ {6}^2 Q_ {9} +324 Q_ {1} Q_ {2}^2 Q_ {3}^3 Q_ {11}\\
 &+ 4 Q_ {6}^3 Q_ {9}+54 Q_ {1} Q_ {3}^2 Q_ {4} Q_ {5} Q_ {11} + 
 324 Q_ {1} Q_ {2} Q_ {3} Q_ {5}^2 Q_ {11} - 
 108 Q_ {1} Q_ {2} Q_ {3}^2 Q_ {6} Q_ {11} - 
 54 Q_ {1} Q_ {5}^2 Q_ {6} Q_ {11}-486 Q_ {2}^2 Q_ {3} Q_ {8} Q_ {11} \\
 & - 81 Q_ {4} Q_ {5} Q_ {8} Q_ {11}+162 Q_ {2} Q_ {6} Q_ {8} Q_ {11} - 4 Q_ {2}^3 Q_ {9} Q_ {13} + 
  Q_ {4}^2 Q_ {9} Q_ {13} - 54 Q_ {1} Q_ {2}^2 Q_ {11} Q_ {13} - 
  12 Q_ {3} Q_ {4} Q_ {5} Q_ {6} Q_ {9}=0,\\
  & 12 Q_ {1}^2 Q_ {2} Q_ {3}^4 Q_ {9} - 
 12 Q_ {1}^2 Q_ {3}^2 Q_ {5}^2 Q_ {9} - 
 5 Q_ {3} Q_ {5}^2 Q_ {7} Q_ {9} - 
 36 Q_ {1} Q_ {2} Q_ {3}^2 Q_ {8} Q_ {9} + 
 18 Q_ {1} Q_ {5}^2 Q_ {8} Q_ {9} + 27 Q_ {2} Q_ {8}^2 Q_ {9} +Q_ {2} Q_ {7} Q_ {9} Q_ {13}\\
 & - 
 3 Q_ {3} Q_ {9}^3-6 Q_ {5}^2 Q_ {7} Q_ {14} + 18 Q_ {9}^2 Q_ {14} - 
  12 Q_ {5} Q_ {12} Q_ {15}=0,\\
  & 12 Q_ {1}^2 (Q_ {2}^3 Q_ {3}^4 - 
 2Q_ {2}^2 Q_ {3}^2 Q_ {5}^2 +  Q_ {2} Q_ {5}^4)-12 Q_ {2}^2 Q_ {3} Q_ {5}^2 Q_ {7} - Q_ {4} Q_ {5}^3 Q_ {7} + 
  3 Q_ {2} Q_ {5}^2 Q_ {6} Q_ {7} - 
  36 Q_ {1} Q_ {2}^3 Q_ {3}^2 Q_ {8} + 
  36 Q_ {1} Q_ {2}^2 Q_ {5}^2 Q_ {8} \\
  & + 27 Q_ {2}^3 Q_ {8}^2 - 
  3 Q_ {4} Q_ {5} Q_ {9}^2 - 3 Q_ {2} Q_ {6} Q_ {9}^2 + 
  Q_ {2}^3 Q_ {7} Q_ {13}=0,\\
  & 12 Q_ {1}^2 (Q_ {2}^2 Q_ {3}^4 Q_ {4} + 12 Q_ {2}^3 Q_ {3}^3 Q_ {5} - 
 2 Q_ {2}^2 Q12 Q_ {1}^2 Q_ {2} Q_ {5}^3 Q_ {6})+(36 Q_ {2}^3 Q_ {3}^2 Q_ {5}  - 
 24 Q_ {2}^2 Q_ {3} Q_ {5} Q_ {6}  - 
 Q_ {4} Q_ {5}^2 Q_ {6}  + 4 Q_ {2} Q_ {5} Q_ {6}^2) Q_ {7}\\
 &+ 
 72 Q_ {1} \left(Q_ {2}^2 Q_ {5} Q_ {6}   -\frac{1}{2} Q_ {2}^2 Q_ {3}^2 Q_ {4}  - 
 3 Q_ {2}^3 Q_ {3} Q_ {5}\right) Q_ {8} + 
 27 Q_ {2}^2 Q_ {4} Q_ {8}^2 - 12 Q_ {2}^2 Q_ {5} Q_ {9}^2-12 Q_ {1}^2 Q_ {2}^3 Q_ {5} Q_ {13}-3 Q_ {4} Q_ {6} Q_ {9}^2\\
 &  +Q_ {2}^2 Q_ {4} Q_ {7} Q_ {13}=0,\\
 & 12 Q_ {1}^2 Q_ {2} Q_ {3}^4 Q_ {4} Q_ {5} + 
 144 Q_ {1}^2 Q_ {2}^2 Q_ {3}^3 Q_ {5}^2 - 
 12 Q_ {1}^2 (Q_ {2}^2 Q_ {3}^4 2 Q_ {2} Q_ {3}^2 Q_ {5}^2) Q_ {6}+36 Q_ {2}^2 Q_ {3}^2 Q_ {5}^2 Q_ {7} - 
 12 Q_ {2} Q_ {3} Q_ {5}^2 Q_ {6} Q_ {7} -27 Q_ {2}^2 Q_ {6} Q_ {8}^2\\
 & + Q_ {5}^2 Q_ {6}^2 Q_ {7}+ 
 36 Q_ {1} (Q_ {2}^2 Q_ {3}^2 Q_ {6} - Q_ {2} Q_ {3}^2 Q_ {4} Q_ {5}+Q_ {2} Q_ {5}^2 Q_ {6} ) Q_ {8} - 
 216 Q_ {1} Q_ {2}^2 Q_ {3} Q_ {5}^2 Q_ {8} + 
 27 Q_ {2} Q_ {4} Q_ {5} Q_ {8}^2 -12 Q_ {2} Q_ {5}^2 Q_ {9}^2\\
 & +3 Q_ {6}^2 Q_ {9}^2 - 12 Q_ {1}^2 Q_ {2}^2 Q_ {5}^2 Q_ {13} + 
  Q_ {2} Q_ {4} Q_ {5} Q_ {7} Q_ {13} - 
  Q_ {2}^2 Q_ {6} Q_ {7} Q_ {13}=0,\\
  & 48 Q_ {1}^2 Q_ {2}^2 (Q_ {3}^2 Q_ {6}^2+10  Q_ {2}^2 Q_ {3}^4)-12 Q_ {1}^2 Q_ {2} Q_ {3}^4 Q_ {4}^2-72 Q_ {1}^2 Q_ {2}^2 Q_ {3}^3 Q_ {4} Q_ {5} - 
 96 Q_ {1}^2 Q_ {2}^3 Q_ {3}^2 Q_ {5}^2-36 Q_ {1}^2 Q_ {2} Q_ {3} Q_ {4} Q_ {5}^3 + 
 48 Q_ {1}^2 Q_ {2}^2 Q_ {5}^4\\
 & -288 Q_ {1}^2 Q_ {2}^3 Q_ {3}^3 Q_ {6} + 
 24 Q_ {1}^2 Q_ {2} Q_ {3}^2 Q_ {4} Q_ {5} Q_ {6} + 
 36 Q_ {1}^2 Q_ {2}^2 Q_ {3} Q_ {5}^2 Q_ {6} - 
 12 Q_ {1}^2 Q_ {2} Q_ {5}^2 Q_ {6}^2+108 Q_ {2}^4 Q_ {3}^3 Q_ {7}-48 Q_ {2}^3 Q_ {3} Q_ {5}^2 Q_ {7}\\
 &  +3 Q_ {3} Q_ {4}^2 Q_ {5}^2 Q_ {7} - 
 4 Q_ {2} Q_ {4} Q_ {5}^3 Q_ {7} - 
 108 Q_ {2}^3 Q_ {3}^2 Q_ {6} Q_ {7} + 
 12 Q_ {2}^2 Q_ {5}^2 Q_ {6} Q_ {7} + 
 36 Q_ {2}^2 Q_ {3} Q_ {6}^2 Q_ {7} - 4 Q_ {2} Q_ {6}^3 Q_ {7} - 
 792 Q_ {1} Q_ {2}^4 Q_ {3}^2 Q_ {8}\\
 &  +36 Q_ {1}\left( Q_ {2} Q_ {3}^2 Q_ {4}^2  + 
 3 Q_ {2}^2 Q_ {3} Q_ {4} Q_ {5}  + 
 4 Q_ {2}^3 Q_ {5}^2  + 
12  Q_ {2}^3 Q_ {3} Q_ {6} Q- 
 Q_ {2} Q_ {4} Q_ {5} Q_ {6}  - 
 2  Q_ {2}^2 Q_ {6}^2\right) Q_ {8}+(108 Q_ {2}^4  - 27 Q_ {2} Q_ {4}^2) Q_ {8}^2\\
 &  - 
  36 Q_ {2}^3 Q_ {3} Q_ {9}^2 + 9 Q_ {3} Q_ {4}^2 Q_ {9}^2 - 
  \left(36 Q_ {1}^2 Q_ {2}^4 Q_ {3}  + 
  12 Q_ {1}^2 Q_ {2}^2 Q_ {4} Q_ {5}  + 
  12 Q_ {1}^2 Q_ {2}^3 Q_ {6} + 4 Q_ {2}^4 Q_ {7}  - 
  Q_ {2} Q_ {4}^2 Q_ {7}\right) Q_ {13}=0.\\
  & 
 \end{split}
\end{equation*}
Any element in the commutant $\mathcal{C}_{\mathcal{U}(\mathfrak{s})}(\mathfrak{n})$ can thus be expressed as the product 
\begin{equation}
P=\prod_{i=1}^{17} Q_i^{a_i},\quad \left(a_1,\dots ,a_{17}\right)\in\mathcal{S},
\end{equation}
where the set $\mathcal{S}$ of constraints consists of the 53 conditions  
\begin{equation*}
\begin{split} 
& a_{10}=0,1,\quad a_{11}=0,1,\quad a_{12}=0,1,\quad a_{14} =0,1,\quad a_{16} =0,1,\quad a_{17}=0,1,\\
& a_2a_{12}=0,\quad a_{5}a_{12}=0,\quad a_2a_{15}=0,\quad a_{2}a_{16}=0,\quad a_{2}a_{17}=0,\quad a_{4}a_{12}=0,\quad a_{4}a_{14}=0,\\
& a_{4}a_{15}=0,\quad a_{4}a_{16}=0,\quad a_{4}a_{17}=0,\quad a_{5}a_{16}=0,\quad a_{5}a_{17}=0,\quad a_{6}a_{14}=0, \quad a_{6}a_{15}=0,\\
& a_{6}a_{16}=0,\quad a_{6}a_{17}=0,\quad a_{7}a_{11}=0,\quad a_{9}a_{10}=0,\quad a_{9}a_{11}=0,\quad 
a_{9}a_{16}=0,\quad a_{9}a_{17}=0,\\
& a_{10}a_{11}=0,\quad a_{10}a_{12}=0,\quad a_{10}a_{14}=0,\quad a_{10}a_{15}=0,\quad a_{10}a_{16}=0,\quad a_{10}a_{17}=0,\quad a_{11}a_{12}=0,\\
& a_{11}a_{14}=0,\quad a_{11}a_{15}=0,\quad a_{11}a_{16}=0,\quad a_{11}a_{17}=0,\quad a_{12}a_{14}=0,\quad a_{12}a_{16}=0,\quad a_{12}a_{17}=0,\\
& a_{14}a_{16}=0,\quad a_{14}a_{17}=0,\quad a_{16}a_{17}=0,\quad  a_{2}a_{6}a_{10}=0,\quad a_{2}a_{10}a_{13}=0,\quad a_{4}a_{10}a_{13}=0,\\
& a_{5}a_{9}a_{12}=0,\quad a_{6}a_{9}a_{12}=0,\quad a_5a_{12}a_{15}=0,\quad   a_{2}a_{4}a_{5}a_6a_7=0,\quad a_{3}a_{4}a_{5}a_6a_7=0,\quad a_{3}a_{4}a_{5}a_6a_9=0,\quad
 \end{split}
\end{equation*}
as well as the requirement $\left(a_1,\dots ,a_{17}\right)\notin \mathbb{Z}\left[v_1,v_2,v_3,v_4\right]$, where
\begin{equation}
 v_1=\left(2,3,4,0^{14}\right),\quad v_2=\left(2,4,4,0^{14}\right),\quad v_3=\left(2,2,4,1,0^{13}\right),\quad  v_4=\left(2,2,3,0,2,0^{12}\right).
\end{equation}
For any admissible $\left(a_1,\dots ,a_{17}\right)\in\mathcal{S}$, the weight $\Lambda=\left[\lambda,\mu\right]$ of the corresponding irreducible representation of $G_2$ is given by 
\begin{equation}
 \begin{split}
\Lambda=\left[2 a_2 + 3 a_4 + a_5 + 2 a_6 + a_9 + 3 a_{10} + 3 a_{11} + 2 a_{12} +
 a_{14} + a_{16}, \right.\\
\left. a_{1} + 2 a_{7} + a_{8} + a_{9} + a_{10} + a_{12} + a_{14} + 2 a_{15} + 
 2 a_{16} + 3 a_{17}\right]
  \end{split}
\end{equation}
with dimension 
\begin{equation}
d_{a_1,\dots a_{17}}=\frac{1}{120}\left(\lambda+1\right)\left(\mu+1\right)\left(\lambda+\mu+2\right)\left(\lambda+2\mu+3\right)\left(\lambda+3\mu+4\right)\left(2\lambda+3\mu+5\right)
\end{equation}
A routine computation shows that the identity (\ref{dek1}) is satisfied for any $p\geq 2$.  
For low values of $p$, the terms in the decomposition (\ref{udek}) are explicitly given by:\footnote{Conventions on representations are adapted to those used in \cite{Pat}.} 
\begin{equation*}
 \begin{split}
p=2:\qquad  \mathcal{U}^{(2)}=& [0,0]\oplus [1,0]\oplus [0,1],\\
p=3:\qquad  \mathcal{U}^{(3)}=&[1,0]\oplus [0,1]\oplus [3,0]\oplus [0,3]\oplus [2,1],\\   
p=4:\qquad  \mathcal{U}^{(4)}=&[0,0]\oplus [2,0]^2\oplus [0,2]^2\oplus [1,1]\oplus [4,0]\oplus [0,4]\oplus [3,1]\oplus [2,2],\\
p=5:\qquad  \mathcal{U}^{(5)}=&[1,0]\oplus [0,1]^2\oplus [1,1]\oplus [1,2]\oplus [2,1]^2\oplus [3,0]^2\oplus [0,3]^2\oplus [3,1]\oplus [3,2]\oplus [2,3]\\
& \oplus [4,1]\oplus [5,0]\oplus [0,5],\\
p=6:\qquad  \mathcal{U}^{(6)}=&[0,0]^2\oplus [2,0]^3\oplus [0,2]^3\oplus [1,1]\oplus [3,0]^2\oplus [2,1]\oplus [1,2]\oplus [4,0]^3\oplus [0,4]^2\oplus [2,2]^2\\
& \oplus [3,1]^2\oplus [1,3]\oplus [3,2]\oplus [6,0]\oplus [5,1]\oplus [4,2]\oplus [3,3]\oplus [2,4]\oplus [0,6].
\end{split}
\end{equation*}
The total number of terms in the decomposition for the values $1\leq p\leq 16$ is 
\begin{equation*}
1, 3, 5, 10, 17, 31, 48, 80, 122, 187, 274, 404, 569, 805, 1106, 1512, 
\end{equation*}
while the number of nonequivalent irreducible representations intervening is 
\begin{equation*}
1, 3, 5, 8, 13, 19, 25, 35, 44, 55, 66, 80, 93, 109, 124, 142. 
\end{equation*}

\medskip
\noindent An interesting property of the commutant $\mathcal{C}_{\mathcal{U}(\mathfrak{s})}(\mathfrak{n})$ of $G_2$ that distinguishes it from previous cases is the existence of several non-Abelian polynomial subalgebras. As commented before, the polynomials $\left\{Q_1,\dots Q_5,Q_{13}\right\}$ generate a six-dimensional Abelian subalgebra $\mathcal{A}_0$. To comment only on some possibilities, we observe that if we extend $\mathcal{A}_0$ adjoining $Q_6$, the commutators show that $\mathcal{A}_1= \left\{Q_1,\dots Q_6,Q_{11},Q_{13}\right\}$ determines an eight-dimensional polynomial algebra. As these eight polynomials are functionally independent, they form an integrity basis for the system (\ref{Rep2}). The subalgebra $\mathcal{A}_1$ suffices to describe the decomposition (\ref{udek}) up to $p=3$, and fails for $p=4$ due to be absence of $Q_7$. If we adjoin the latter element, the resulting polynomial algebra is easily seen to be isomorphic to the whole commutant. However, adjoining $Q_{10}$ to $\mathcal{A}_1$ leads to a nine-dimensional subalgebra $\mathcal{A}_2$ that is maximal, as the adjunction of any other element again generates the commutant $\mathcal{C}_{\mathcal{U}(\mathfrak{s})}(\mathfrak{n})$. On the other hand, the elements $\left\{Q_1,Q_2,Q_4,Q_{10}\right\}$ themselves generate a minimal non-Abelian polynomial algebra.

\section{Conclusions}

The problem of determining the decomposition of the universal enveloping algebra of a simple complex Lie algebra $\mathfrak{s}$ has been revisited combining both the purely algebraic and the analytical approach. It has been observed that determining the polynomials that commute with the elements associated to the simple roots of $\mathfrak{s}$ is formally equivalent to computing the subgroup scalars for the reduction chain $\mathfrak{n}\subset \mathfrak{s}$, where $\mathfrak{n}$ is the maximal nilpotent ideal of the Borel subalgebra of $\mathfrak{s}$. This fact allows us to deduce that the commutant $\mathcal{C}_{\mathcal{U}(\mathfrak{s})}(\mathfrak{n})$ always possesses a maximal Abelian subalgebra, the dimension of which is completely determined by $\mathfrak{n}$. With the analyticalapproach, the commutant can be constructed successively starting from an integrity basis for the system of PDEs associated to the reduction chain. In this context, the question whether any integrity basis forms a non-Abelian polynomial algebra has been shown to be false in general, although for any of the cases considered, a polynomial subalgebra $\mathcal{A}$ of the commutant $\mathcal{C}_{\mathcal{U}(\mathfrak{s})}(\mathfrak{n})$ and whose elements form an integrity basis has been shown to exist. 

\medskip
The case of rank two simple Lie algebra has been analyzed in detail. For the classical algebras $A_2$ and $B_2$, that have already been considered in the literature (see \cite{Fla,Bur1,Bur2}), additional properties of the commutant have been described. The case of the exceptional Lie algebra $G_2$, which has been solved for the first time, is much more complicated. The first notable difference with the classical algebras is that number of linearly polynomials required for the decomposition (\ref{udek}) is more than twice the cardinal of an integrity basis. Besides the computational difficulties due to the degree and length of the polynomials involved, for $G_2$ the number of algebraic dependence relations is unexpectedly high, namely $57$, in contrast with the classical algebras, where the number of relations is low. It has been further shown that, among the three classical Lie algebras of rank two, for $A_2$ and $B_2$ the polynomial subalgebra $\mathcal{A}$ is actually maximal of codimension one, for $D_2$ it does not exist as the commutant is itself Abelian, while for $G_2$ the subalgebra is of codimension nine and not maximal, as it can be extended nontrivially.  

\bigskip
In view of the analogies with the so-called internal label problem \cite{Shar}, we may ask whether the procedure can be adapted to other reduction chains $\mathfrak{s}^{\prime}\subset\mathfrak{s}$ of semisimple Lie algebras in order to construct polynomial algebras. It shall however taken into account that polynomials $P$ in the enveloping algebra of  $\mathfrak{s}$ commuting with the generators of $\mathfrak{s}^{\prime}$ are not suitable to describe the decomposition (\ref{udek}), as the eigenvalue of $P$ with respect to the Cartan subalgebra of $\mathfrak{s}^{\prime}$ is always zero. Nonetheless, the ansatz can be of use for the analysis of integrable systems \cite{Ian,cor} endowed with some internal symmetry, as well as for (quasi-)exactly solvable (QES) systems \cite{gab,cam21}. In \cite{cam21}, the case of $\mathfrak{gl}(3)$ was exploited to provide new algebraic (QES) systems, while in \cite{cor},  $\mathfrak{su}(3)$ was used to provide ab algebraic setting for a superintegrable Hamiltonian on the sphere. This points out how the results of this paper may be as well of interest for different areas of mathematical physics, as well as it may also reveal to be of interest in the description of wider types of representations of simple Lie algebras \cite{Hum}. 

\medskip
As an illustrating example, let us consider the simple Lie algebra $A_2$ in the basis (\ref{TA2}) and the regular subalgebra $\mathfrak{sl}(2,\mathbb{C})$ generated by $H_1,E_{12},E_{21}$. In this case, the generators are no more related to highest weight vectors of $A_2$, but formally, the commutant of the subalgebra $\mathfrak{sl}(2,\mathbb{C})$ in $\mathcal{U}(\mathfrak{sl}(3,\mathbb{C}))$ has also a polynomial structure. An integrity basis for the system (\ref{Rep2}) associated to the subalgebra generators contains five elements, as can be easily verified. A short computation shows that there exist six linearly independent elements that span the commutant, given by the polynomials 
\begin{equation}
\begin{split}
P_1= &H_1+2H_2,\quad P_2= E_{13}E_{31}+E_{23}E_{32}-H_2-\frac{1}{2}H_1,
\quad P_3=H_1^2+4E_{12}E_{21}-2H_1,\\
P_4=& E_{12}E_{31}^2-E_{21}E_{32}^2-H_1E_{31}E_{32},\quad P_5=E_{12}E_{23}E_{31}+E_{13}E_{21}E_{32}+H_1E_{13}E_{31}-H_2E_{23}E_{32}\\
+& E_{23}E_{32}-E_{12}E_{21}-H_2^2-H_1H_2-\frac{1}{2}H_1^2-H_2,\quad  P_6= E_{12}
E_{23}^2-E_{13}E_{21}^2+H_1E_{13}E_{23}-2E_{13}E_{23}.
\end{split}
\end{equation}
The six operators generate a polynomial algebra with nonvanishing commutators  
\begin{equation}
\begin{split}
\left[P_1,P_4\right]=& -6P_4,\quad \left[P_1,P_6\right]= 6P_6,\quad \left[P_2,P_3\right]=3P_4+P_1P_4,\quad \left[P_2,P_6\right]=3P_6-P_1P_6,   \\
\left[P_4,P_5\right]=& \left(-\frac{1}{2}  P_1^2   - 5  P_1     + 4  P_2  - 
\frac{1}{2} P_3-12\right) P_4,\quad \left[P_4,P_6\right]=P_1 P_2P_3 - P_1(P_2^2 -P_2)  +  
( 2  P_2- P_3)P_5\\
& - \frac{1}{4}P_1P_3,\quad 
\left[P_5,P_6\right]= 4  P_2P_6-\frac{1}{2}(P_1^2+P_3)P_6 +5P_1P_6-12P_6.
 \end{split}
\end{equation}
The polynomials are further algebraically dependent through the relation
\begin{equation}
\begin{split}
P_1^2 P_2^2 & - P_2^2 P_3 - 4 P_1 P_2P_5+ 2 P_1 P_2 P_3  - 2 P_1 P_2^2 - P_1^2 P_2 - \frac{1}{4}P_1^2 P_3- 4 P_4P_6+ 4 P_5^2\\
&+ 2\left( P_1 + 2 P_2 -  P_3\right)P_5 +2P_1P_2 -P_2P_3+\frac{1}{4}\left(P_3-2P_1\right)P_3=0,
\end{split}
\end{equation}
showing that any  element in the commutant can be written as 
\begin{equation}
P= P_1^{a_1}P_2^{a_2}P_3^{a_3}P_4^{a_4}P_5^{a_5}P_6^{a_6},\quad a_i\in\mathbb{N}\cup{0},\quad a_5=0,1.
\end{equation}
We observe that the weight of the polynomials with respect to the Cartan subalgebra of $\mathfrak{sl}(3,\mathbb{C})$ is (0,0) for $P_1,P_2,P_3$ and $P_5$, while $P_4$ has weight $(0,3)$ and $P_6$ has weight $(0,-3)$. 

\medskip

\section*{Acknowledgement}
RCS was   supported by
the research grant PID2019-106802GB-I00/AEI/10.13039/501100011033 (AEI/ FEDER, UE).
IM was supported by by Australian Research Council Future Fellowship FT180100099.

{}

\end{document}